\documentclass{aa}  
\usepackage{graphicx}
\usepackage{txfonts}
\begin{document} 

   \title{The twisted jets and magnetic fields of the extended
radio galaxy 4C\,70.19\thanks{Based on observations with the 100-m telescope at Effelsberg
operated by the Max-Planck-Institut f\"ur Radioastronomie (MPIfR) on behalf of the Max-Planck-Gesellschaft.}
    }
\author{M. We\.zgowiec\inst{1}
\and M. Jamrozy\inst{1}
\and K. T. Chy\.zy\inst{1}
\and M. J. Hardcastle\inst{2}
\and A. Ku\'zmicz\inst{1}
\and G. Heald\inst{3}
\and T. W. Shimwell\inst{4}}
\institute{
Obserwatorium Astronomiczne Uniwersytetu Jagiello\'nskiego, ul. Orla 171, 30-244 Krak\'ow, Poland, \\
\email{markmet@oa.uj.edu.pl}
\and Centre for Astrophysics Research, Department of Physics, Astronomy and Mathematics, University of Hertfordshire, College Lane, Hatfield AL10 9AB, UK
\and CSIRO Astronomy and Space Science, PO Box 1130, Bentley WA 6102, Australia
\and ASTRON, the Netherlands Institute for Radio Astronomy, Postbus 2, 7990 AA, Dwingeloo, The Netherlands}
\offprints{M. We\.zgowiec}
\date{Received; accepted date}

\titlerunning{The twisted jets and magnetic fields of 4C\,70.19}
\authorrunning{M. We\.zgowiec et al.}

% \abstract{}{}{}{}{} 
% 5 {} token are mandatory
 
  \abstract
   {
 The appearance of the jets and lobes of some radio galaxies makes it difficult to assign them to a known class of objects. 
This is often due to the activity of the central engine and/or interactions with the environment, as well as projection effects.
   }
   {
We analyse the radio data for an apparently asymmetric radio source 4C\,70.19, which is associated with the giant elliptical galaxy NGC\,6048.
The source shows distorted radio jets and lobes, one of which bends by 180$\degr$. The aim of our study is to explain the nature of the observed distortions.
   }
   {
We used LOFAR, Effelsberg, and VLA radio data in a wide range of frequencies. At high frequencies, we also used radio polarimetry
to study the properties of the magnetic fields. Additionally, we made use of optical, infra-red, and X-ray data. 
   }
   {
Polarisation data suggest shearing of the magnetic fields at points where the jets bend.
The low-frequency LOFAR map at 145\,MHz, as well as the sensitive single-dish Effelsberg map at 8.35\,GHz, reveal previously undetected diffuse emission
around the source. The rotation measure (RM) derived from the polarimetric data allowed us
to estimate the density of the medium surrounding the source, which agrees with typical densities of the intergalactic medium or the outer
parts of insterstellar halos.
   }
   {
We propose that the southern jet is bent in the same manner as the northern one,
but that it is inclined to the sky plane. Both these bends are likely caused by the orbital motion within the galaxy group,
as well as interactions with the intergalactic medium. Our analyses suggest that, despite its complex morphology,
4C\,70.19 seems to be intrinsically symmetric with a physical extent of up to 600\,kpc, and that the diffuse emission
detected in our high-sensitivity maps is related to radio plumes that are expanding behind the source.
   }


   \maketitle
%
%________________________________________________________________

\section{Introduction}
\label{intro}

Radio galaxies (RGs) are known to possess an active galactic nucleus (AGN), which is responsible for the production of jets and lobes,
often extending far from the host galaxy to distances of
tens to hundreds of kiloparsecs (kpc) or even up to several megaparsecs (Mpc), as reported by, for example, \citet{machalski08}, \citet{kuzmicz18}, 
or \citet{dabhade20}.
It is clear that those jets and lobes must interact with the surrounding intergalactic medium (IGM).
In the case of objects that form part of a larger system (galaxy groups or clusters), the interactions also include the gravitational influence
experienced by the source along its orbit \citep[e.g.][]{vallee81,horellou18}. Simultaneously,
the same kinds of interactions can stimulate AGN activity.
The complex morphology of the radio jets and lobes of RGs often suggests an episodic activity of the central source or the influence
of the orbital motion within their group or cluster environment. Recurrent activity has been reported in a number of sources
 \citep[e.g.][and references therein]{saikia09,nandi12,kuzmicz17,mahatma19}.

In this paper, we present the results of our investigation of the radio galaxy 4C\,70.19, 
the radio jets of which appear to have a distorted morphology, including an unusual, 
asymmetric structure with a 'hook-like' jet on one side \citep{lara01}.
The source does not show prominent hot spots and was recognised by \citet{lara01} as an FRI source \citep{fanaroff74}.
However, the brightenings of the outer parts of the source, which form lobe-like structures, require further investigation.
To analyse and explain the morphology of 4C\,70.19, we gathered data over a wide range of radio frequencies, from
LOw-Frequency ARray \citep[LOFAR;][]{vanhaarlem13} data at 145\,MHz to sensitive single-dish Effelsberg observations at 8.35\,GHz.
The high-frequency radio data enable polarisation studies, which allowed us to examine the properties of the magnetic fields throughout
the entire radio source. Sensitive single-dish observations above 1.4\,GHz allowed us to study the evolution of relativistic electrons in 4C\,70.19,
which led to reliable determinations of the spectral index profiles and both the dynamical and synchrotron age.
To further study the intergalactic environment of 4C\,70.19, we also analysed the vicinity of this source using archival X-ray,
optical, and infrared data.

Throughout this paper we use a cosmological model in which H$_{\rm 0}$ = 71\,km\,s$^{-1}$\,Mpc$^{-1}$, $\Omega_{\rm M}$ = 0.27,
and $\Omega_\Lambda$ = 0.73 \citep{spergel03}. This translates
the redshift of the source, of $z=0.0257$, to a co-moving distance of 108\,Mpc (1 arcminute corresponds to 30\,kpc).

4C\,70.19 is associated with the bright (R $\simeq$ 12.7 mag) galaxy NGC\,6048
(also known as UGC\,10124),
located at Right Ascension (RA) = 15$^{\rm h}$57$^{\rm m}$30\fs2, (Dec.) = +70\degr41$^\prime$21\farcs0 (J2000.0).
In the optical spectrum of NGC\,6048, only absorption lines are present \citep{stickel93},
which suggests that the central AGN is accreting at a low level,
which in turn is typical for an FRI source \citep{hinelongair79}.
\citet{schombert15} classified NGC\,6048 as a D-class elliptical galaxy, being the result of a recent
equal-mass dry merger, that is, of two gas-poor galaxies (no energy dissipation via the gaseous component).

The radio structure of 4C\,70.19 has a large angular extent of about 11\arcmin\ in the N-S direction,
which corresponds to a projected linear size of about 330\,kpc.
The large angular size enables studies of the physical conditions throughout the source.
The extended radio morphology of 4C\,70.19, with a bright core, jets, and radio plumes, resembles that
of an FRI-type RG, and the source has a total 1400 MHz radio power of $\log$(P) = 24.5\,W\,Hz$^{-1}$,
which is below the canonical FRI/FRII luminosity break.
A closer look, however, reveals some unusual features.
Previous radio observations with NVSS show that the northern radio jet is bent by around 180$\degr$ and then
extends into a plume-like structure.
On the contrary, the southern jet, which has a relatively typical morphology, is slightly deviated
from the main axis of the radio source.
No prominent compact hot spots have been detected in the high-frequency high-resolution maps \citep[e.g.][]{lara01}.

Throughout the paper we use two names for our studied source: we use 4C\,70.19 whenever radio data are analysed, 
and NGC\,6048 when discussing other data
referring to the host galaxy.

\section{Observations and data reduction}
\label{obsred}

In this paper, we use LOFAR survey data and Effelsberg observations, as well as other
archival radio data. To reveal the nature of 4C\,70.19, our study also uses infrared maps
from the Wide-field Infrared Survey Explorer \citep[{\sc WISE};][]{wright10}
and X-ray data from the ROentgen SATellite \citep[{\sc ROSAT};][]{truemper82} High-Resolution Imager \citep[{\sc HRI};][]{pfeffermann87}.

\subsection{Radio data}

\subsubsection{LOFAR}

The inspection of an early LOFAR Multifrequency Snapshot Sky Survey image \citep[MSSS;][]{heald15} revealed that the morphology
of 4C\,70.19 is significantly different at low frequencies, showing a typical straight northern radio lobe.
Its bend, however, known from the radio maps at higher frequencies, was confirmed by the deeper LOFAR Two-metre Sky Survey \citep[LoTSS;][]{shimwell17} images.
The LoTSS data were first processed using a direction independent calibration pipeline \citep[see e.g.][]{vanweeren16,williams16,degasperin19}.
After this the data were processed by the latest version of DDF-pipeline\footnote{https://github.com/mhardcastle/ddf-pipeline},
which makes use of kMS \citep{tasse14,smirnov15} for calibration and DDFacet \citep{tasse18} for imaging.
This pipeline is described in detail in \citet{shimwell19} and \citet{tasse21}.
We used the data from the LoTSS pointing P232+70 (proposal code LC10\_013). Our target galaxy is 2$\fdg$3 from the pointing centre
(59\% of the peak primary beam response).
In this paper, we present the radio map of 4C\,70.19 in the 120-168\,MHz band with a central frequency of 145\,MHz (Fig.~\ref{2m}).
The sensitivity r.m.s level in the vicinity of the source is 0.2\,mJy\,beam$^{-1}$ (see Table~\ref{maps} for details).

\subsubsection{Effelsberg}

We performed high-frequency radio polarimetry of 4C\,70.19.
The observations at 4.85 and 8.35\,GHz were performed in May 2014 (project number 117-13) and
at 2.67\,GHz in November 2014, using the 100-m Effelsberg radio
telescope of the Max-Planck-Institut f\"ur Radioastronomie\footnote{http://www.mpifr-bonn.mpg.de} (MPIfR)
in Bonn. At 4.85\,GHz the observations were performed using the two-horn system (with a horn separation of 8\arcmin)
at the secondary focus \citep[see][]{gioia82}.
Each horn was equipped with two total power receivers and an IF
polarimeter resulting in four channels containing the Stokes
parameters I (two channels), Q and U. The telescope pointing was corrected by
performing cross-scans of a bright point source close to the
observed RG. The flux density scale was established by
mapping the point source 3C\,286 and aligning its flux with the scale of \citet{baars77}.
The same source was also used for polarisation calibration.
To map the source at 4.85\,GHz with the dual-beam receiver, scans in the azimuth-elevation frame were obtained.
At 2.67 and 8.35\,GHz (single horn receivers), maps were obtained by alternately scanning along the R.A. and
Dec. directions. The data reduction was performed using the NOD2 data reduction package
\citep[see][for details]{wezgowiec07}. In the case of data at 2.67\,GHz
we could not use the broadband channel, as some of the frequencies were heavily affected by the RFI.
Instead, two distorted channels were excluded and the remaining data from six channels were used.

\subsubsection{Archival radio data}
\label{archivalradio}
This paper also uses archival VLA data at 1.43 and 4.85\,GHz, which were reduced
using the {\sc AIPS} package\footnote{http://aips.nrao.edu}. The WENSS map at 327\,MHz was also used in its original version
to provide data for spectral analysis of the source.

At 1.43\,GHz we used VLA C-configuration data from project AL442. For the flux density scale we used the radio
source 3C\,138, which also served for the calibration of the polarisation angle. The phase calibration was performed with
the use of the radio sources J1435+760 and J1748+700.

At 4.85\,GHz two data sets in C (project BT012B) and D (project AW269) configurations of the VLA were used.
For both data sets the flux density scale and polarisation angle calibration was performed with the use of the radio source 3C\,286.
The phase calibration was done using the radio sources J1531+721 and J1642+689 for C and D configurations, respectively. To ensure both
high resolution and sensitivity to large-scale radio emission, the calibrated data from both configurations were combined prior to imaging
with the AIPS task {\sc dbcon}.

The image deconvolution from the calibrated VLA data was performed using the Clark {\sc clean} algorithm \citep{clark80}
and resulted in elliptical beams of the clean maps. Each beam was later convolved to a circular one. Because of the large angular
extent of 4C\,70.19, the observations at 4.85\,GHz, due to the largest angular scale of only around 4\arcmin\
at both C and D configurations\footnote{https://science.nrao.edu/facilities/vla/docs/manuals/oss/\\performance/resolution},
suffer from a lack of information about the large-scale emission. Therefore, we show these data only in combination with
the corresponding single-dish Effelsberg maps.

The combination of maps was performed in the spatial frequency domain image plane using the AIPS task {\sc imerg}.
The task Fourier-transforms the images to the (u,v) plane, normalises amplitudes within the annulus of overlap in the (u,v) plane and
back transforms the combined data to the image plane. The normalisation uses the ratio of the resolutions of the input images.
To verify the quality of the combined maps, we compared the flux densities in the Effelsberg maps with those in the combined maps and
found that the fluxes agree within 3\%.

\subsubsection{Polarisation radio data}
\label{polar}

From the Effelsberg and the VLA radio data we obtained maps of the polarised emission, as well as the
polarisation angle maps. These maps were then used to plot the so-called B-vectors, which show the orientation of the magnetic field
(polarisation angle of the E-vectors rotated by 90\degr). The length of the B-vectors is presented as proportional to the polarised intensity.
The maps of the polarisation angle were also used to produce maps of the RM, which provide information about the magnetic field
along the line of sight and help to estimate the density of the IGM around 4C\,70.19 (see Sect.~\ref{magfields} for details).

\subsection{Other data}

The available pipeline products of the ROSAT HRI pointed observations (Dataset ID rh701836N00) of NGC\,6048 were used
to obtain the HRI map of the X-ray emission and the countrate of the core region. The latter allowed us to estimate the luminosity
of the central source of 4C\,70.19.
The HRI map was adaptively smoothed in the full energy range of the instrument using the {\sc asmooth} task from the
XMM-Newton Science Analysis System \citep[SAS;][]{gabriel04} package.

From the WISE archives maps in all 4 energy bands covered by the telescope (W1 -- 3.4\,$\mu$m, W2 -- 4.6\,$\mu$m, W3 -- 12\,$\mu$m, and W4 -- 22\,$\mu$m)
were acquired to trace the dust emission in the main body of NGC\,6048.

We also made use of several optical catalogues, as well as the Pan-STARRS data \citep{flewelling20}, to obtain information about the group environment
of NGC\,6048, including its closest companions (see Sect.~\ref{multi}).

\section{Results}
\label{results}

\subsection{Radio maps}
\label{radiomaps}

Sensitive radio maps of 4C\,70.19 in a wide range of frequencies and resolutions (see Table~\ref{maps})
allowed us to study the morphology of both radio jets (also within the host galaxy) and of the radio envelope.

\begin{table}[ht]
\caption{\label{maps}Radio maps and fluxes of 4C\,70.19}
\centering
\begin{tabular}{rlrlr@{$\pm$}l}
\hline\hline
Freq.   & Telescope     & Beam   & r.m.s.        &\multicolumn{2}{c}{Total}\\
        &               & size   & [mJy          &\multicolumn{2}{c}{flux}\\
\,[MHz] &               &[\arcsec]&beam$^{-1}$]&\multicolumn{2}{c}{[Jy]}\\
\hline
145     & LOFAR         & 6      & 0.20         &9.29&0.47      \\
327     & WSRT          & 60     & 7.00         &5.15&0.26      \\
1430    & VLA           & 34     & 0.37         &2.60&0.13      \\
2674    & Effelsberg    & 264    & 3.50         &1.55&0.08      \\
4850    & VLA+Effelsberg& 15     & 0.29         &1.03&0.05      \\
8350    & Effelsberg    & 82     & 0.31         &0.76&0.04      \\
\hline
\end{tabular}
\end{table}

\begin{figure*}
	\sidecaption
        \includegraphics[width=12cm,clip]{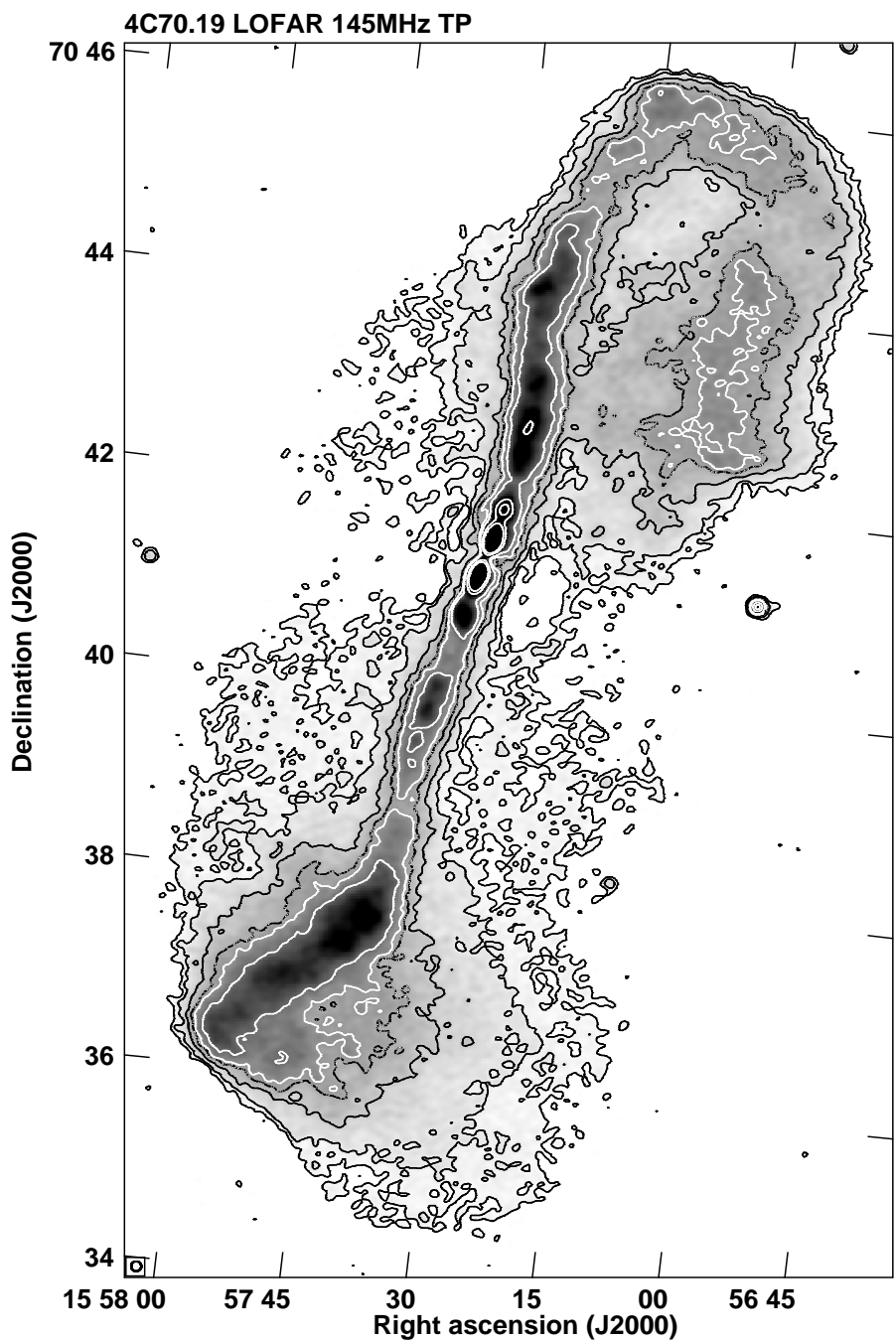}
\caption{
        LOFAR map of 4C\,70.19 at 145\,MHz with contours of (3, 6, 12, 18, 22, 30, 64, and 96)\,$\times$\,0.2\,mJy\,beam$^{-1}$.
        The beam size of 6\arcsec\ is shown in the bottom left corner of the image. The compact object below the northern plume is an 
        unrelated background source.
        }
\label{2m}
\end{figure*}

\begin{figure}
        \resizebox{\hsize}{!}{\includegraphics[clip]{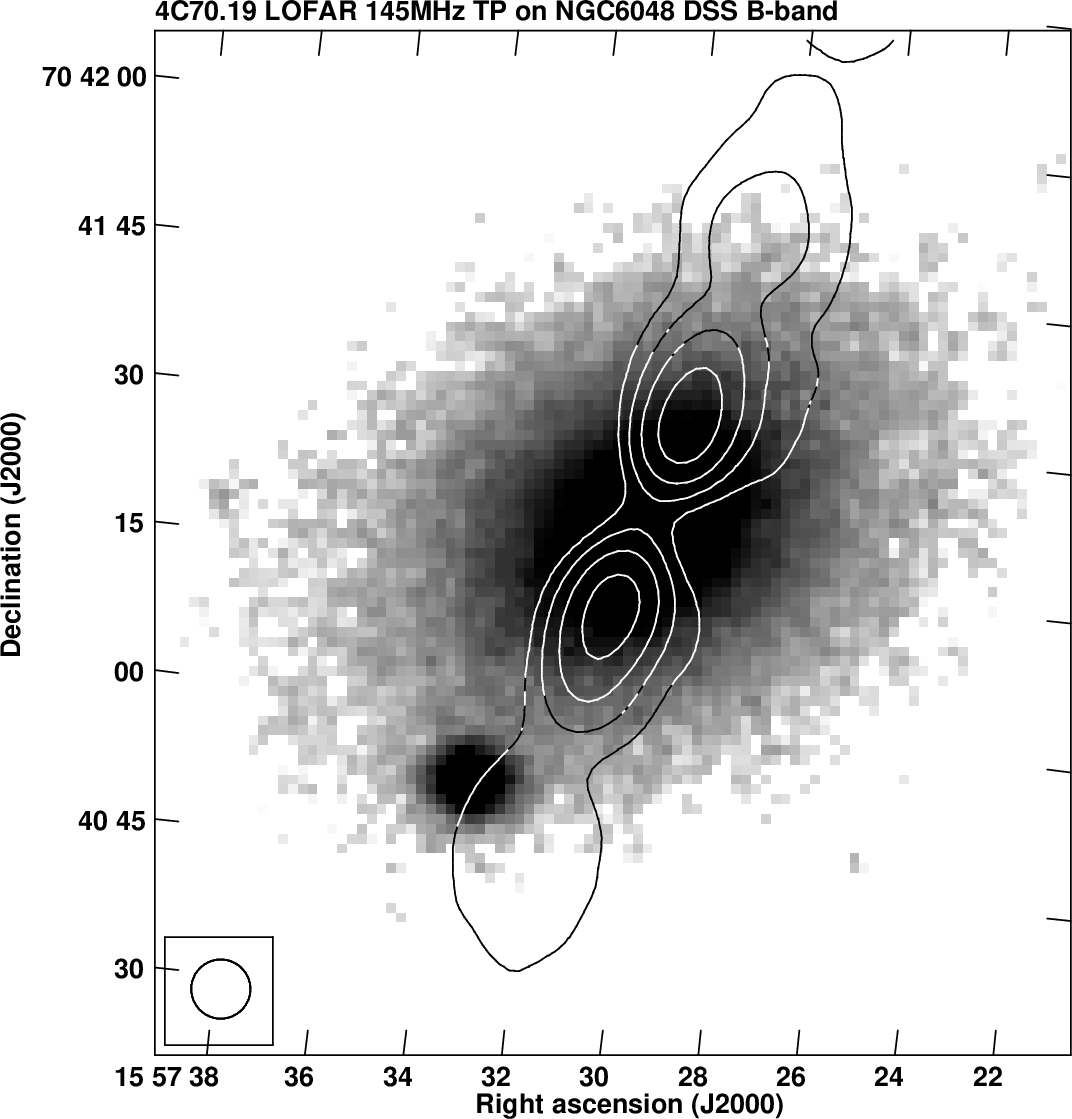}}
\caption{LOFAR map of radio total power at 145\,MHz of the inner jet structures overlaid on the DSS B-band image.
        The contours are (32, 64, 128, and 256)\,$\times$\,0.2\,mJy\,beam$^{-1}$.
        The resolution of the map is 6\arcsec.}
\label{inner}
\end{figure}

The capabilities of the LOFAR telescope allowed us to detect both the diffuse extended emission around the source and
fine details of the radio jets, also within the host galaxy (Figs.~\ref{2m} and~\ref{inner}).
Unlike in the high frequency maps of \citet{lara01}, in the 145\,MHz LOFAR map the central source of 4C\,70.19 is not visible.
The two distinct inner structures that are likely bases of the jets, also detected by \citet{lara01}, are well resolved.
The jets extending further out are not coaxial and change their direction. The high resolution of the map confirms that
at low frequencies the morphology of the source is very similar to that found in the high frequency maps and that the northern plume
is formed by a jet, which changes its direction by 180\degr. The southern jet also bends significantly towards southeast and
the visible kink at its end suggests that also in the south a significant change in the direction of the jet is observed.
In particular, more diffuse emission extends to the west, which may mark the southern plume, oriented not directly in the sky plane.
Around the bright jets some diffuse large-scale emission can be traced. Its patchy appearance likely originates
from the low sensitivity to faint large-scale structures of the small beam of the
observations. The large-scale diffuse radio emission at low frequencies will be discussed in Sect.~\ref{morph}.

The zoomed-in central part of the LOFAR map overlaid on the DSS\footnote{https://archive.stsci.edu/cgi-bin/dss\_form}
optical image of NGC\,6048 (Fig.~\ref{inner}) shows that the bases of the jets
reside within the main body of the galaxy, slightly extending into the halo, where the secondary brightenings are present.

\begin{figure}
\resizebox{\hsize}{!}{\includegraphics[clip]{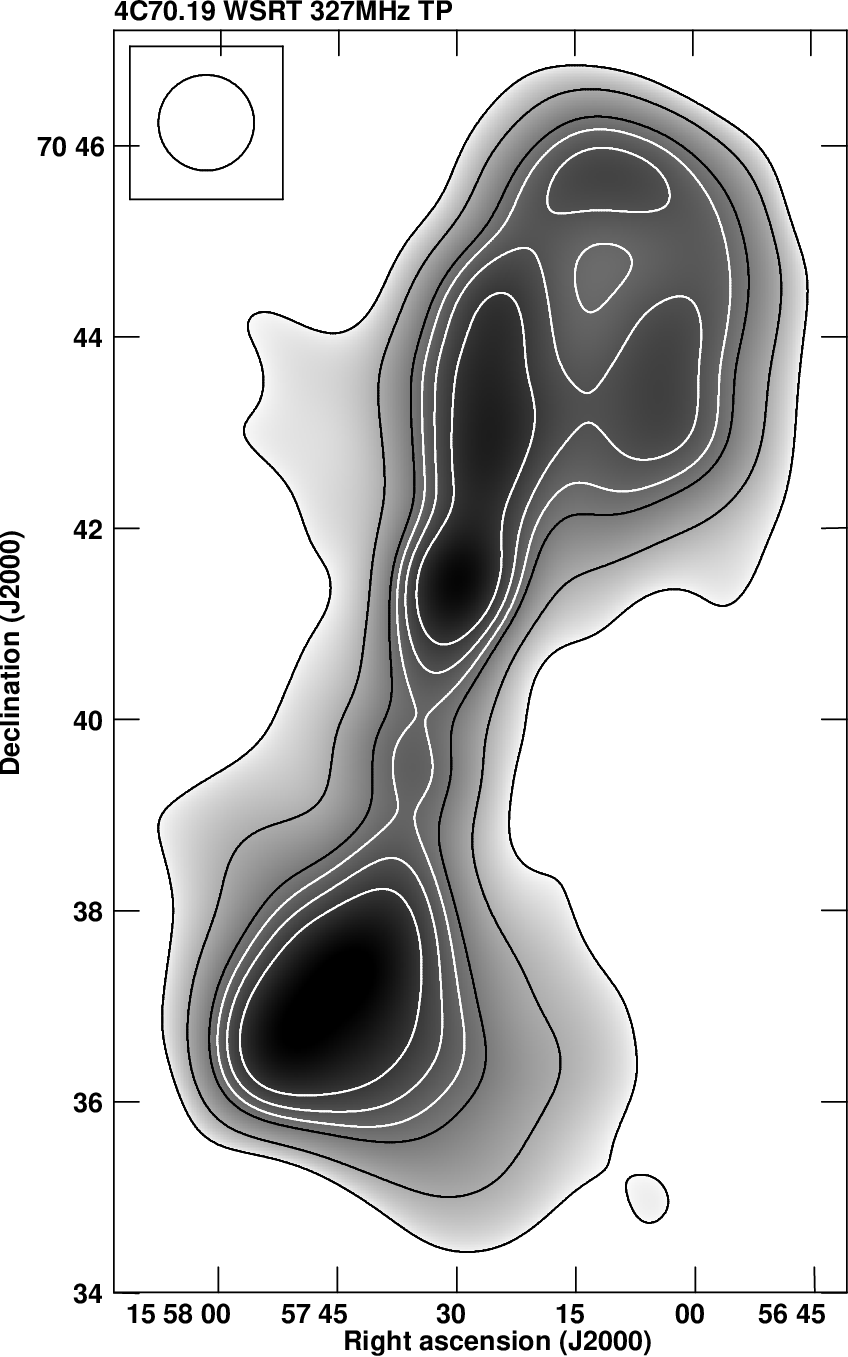}}
        \caption{WENSS map of 4C\,70.19 at 327\,MHz with contours of (3, 8, 16, 20, 25, and 32)\,$\times$\,7\,mJy\,beam$^{-1}$. The beam size of 60\arcsec\ is
presented in the upper left corner of the image.
}
\label{wenss}
\end{figure}

The WSRT map at 1\arcmin\ resolution at 327\,MHz from the WENSS survey (Fig.~\ref{wenss}) shows the global
morphology of the source. The hook-like shape of the northern jet is only slightly resolved.
On both sides of the source the emission surrounding the jet regions does not extend farther than that visible in the LOFAR map.

\begin{figure}
        \resizebox{\hsize}{!}{\includegraphics[clip]{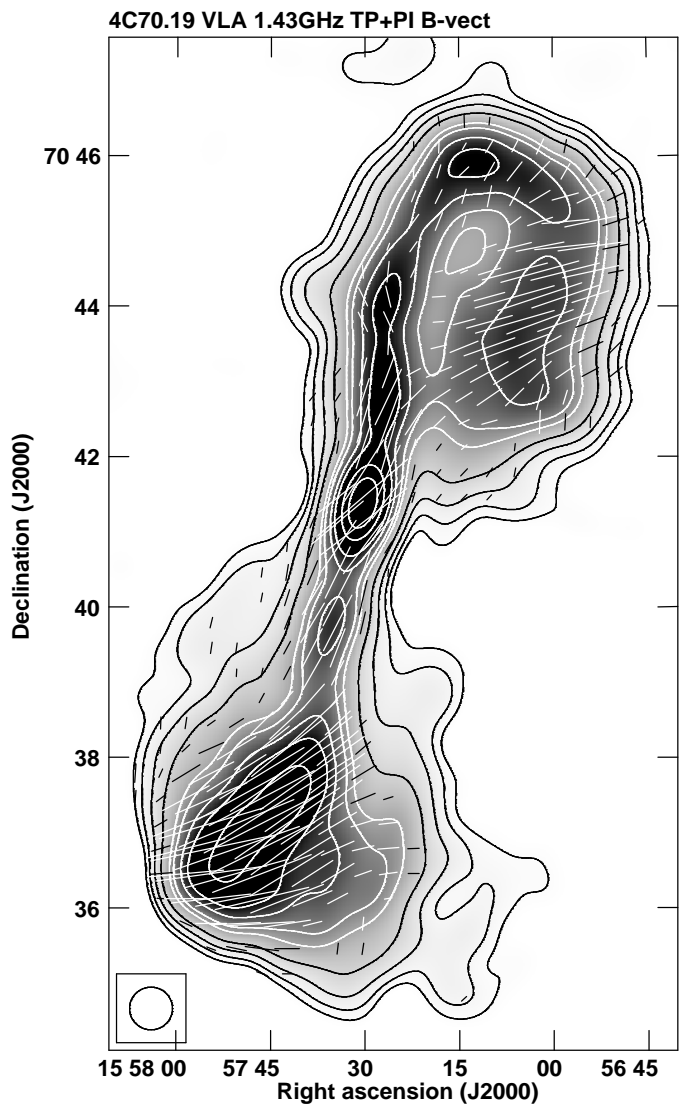}}
\caption{
        VLA map of 4C\,70.19 at 1.43\,GHz. The contours are (3, 6, 12, 24, 48, 64, 96, 128, 192, and 256)\,$\times$\,0.37\,mJy\,beam$^{-1}$.
        The lines show the orientation of the magnetic field and a length of 1\arcmin\ corresponds to polarised emission of 10\,mJy\,beam$^{-1}$.
        The beam size of 34\arcsec\ is presented in the bottom left corner of the image.
        }
\label{21cm}
\end{figure}

The core region of 4C\,70.19 and the bend of the northern jet becomes resolved in the VLA map at 1.43\,GHz (Fig.~\ref{21cm}).
Also the elongated area of higher intensity at the end of the southern jet is distinctly visible. The extensions of the weaker emission visible
to the southwest and northeast from this area correspond very well with the high resolution map at 145\,MHz.
Despite Faraday rotation, expected to be high at this frequency, the B-vectors are parallel to the jet entering the northern lobe and
almost perpendicular to it at the tip of the hook. In this sense the southern lobe is different and the systematic change in the directions
of B-vectors is not observed.

\begin{figure}
\resizebox{\hsize}{!}{\includegraphics[clip]{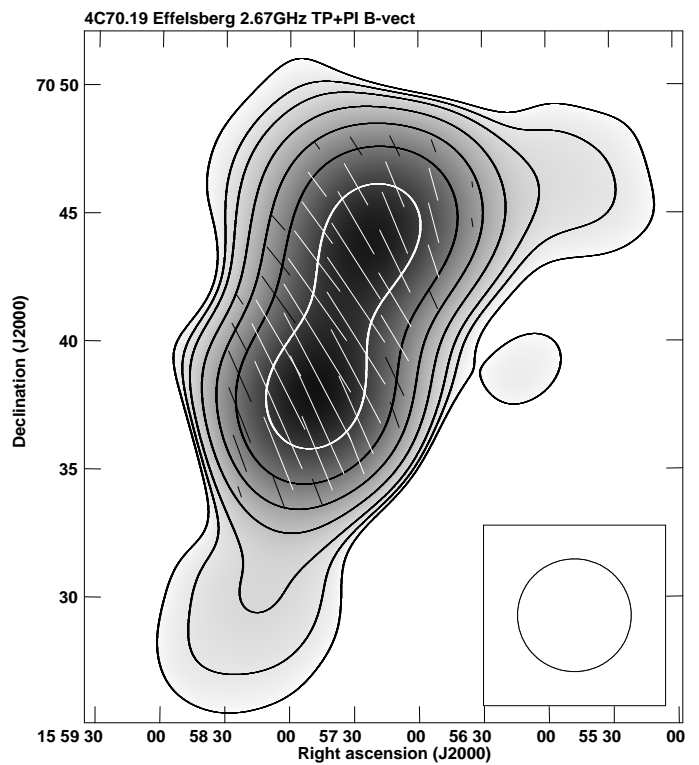}}
        \caption{Effelsberg map of 4C\,70.19 at 2.67\,GHz. The contours are (3, 5, 8, 16, 32, 64, and 128)\,$\times$\,3.5\,mJy\,beam$^{-1}$.
The lines show the orientation of the magnetic field and a length of 1\arcmin\ corresponds to polarised emission of 10\,mJy\,beam$^{-1}$.
The beam size of 264\arcsec\ is presented in the bottom right corner of the image.
}
\label{11cm}
\end{figure}

The low resolution map which we obtained with the Effelsberg telescope at 2.67\,GHz (Fig.~\ref{11cm}) shows
just the overall shape and the main orientation of 4C\,70.19 and was used
only to measure the total flux of the source. The extensions to the south and west are due to background radio sources
and were not included in the flux measurements (see Sect.~\ref{spixdisc}).
Due to the beam depolarisation and Faraday rotation effects, only the approximate (northeast-southwest)
direction of the B-vectors could be obtained.

\begin{figure}
        \resizebox{\hsize}{!}{\includegraphics[clip]{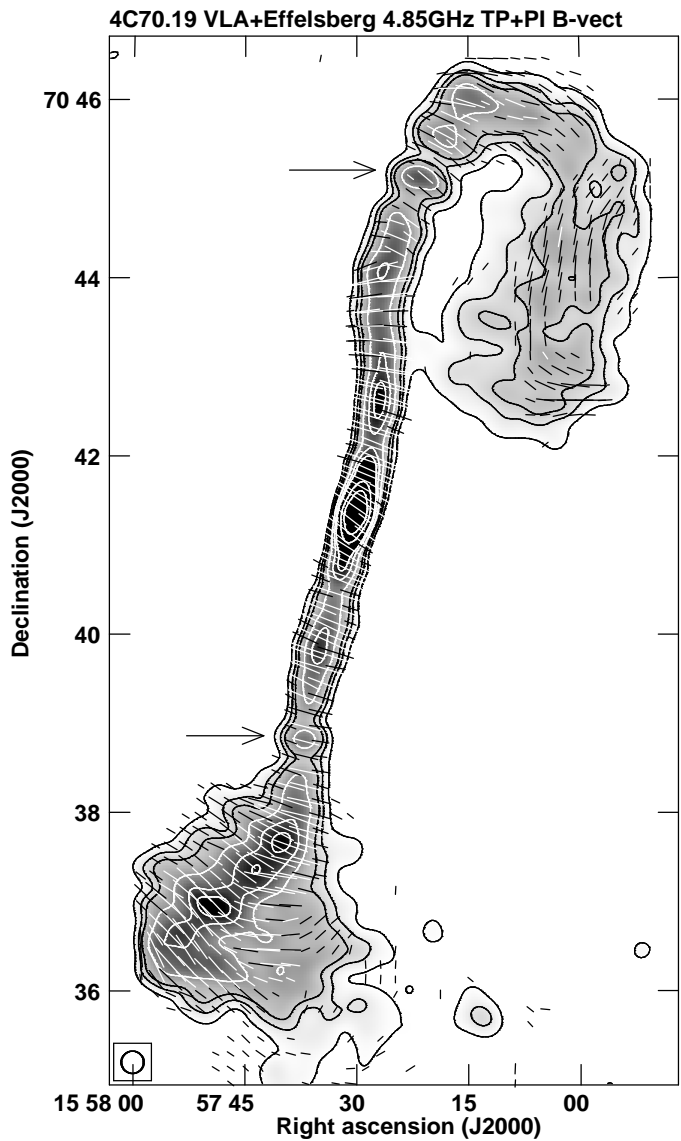}}
\caption{
        Merged map of 4C\,70.19 at 4.85\,GHz produced from the Effelsberg single-dish map and the VLA interferometer maps in C and D configurations.
        The contours are (3, 6, 9, 16, 25, 32, 64, and 96)\,$\times$\,0.29\,mJy\,beam$^{-1}$.
        The lines show the orientation of the magnetic field and a length of 1\arcmin corresponds to polarised emission of 3.3\,mJy\,beam$^{-1}$.
        The beam size of 15\arcsec\ is presented in the bottom left corner of the image. The two arrows point at bright knots of the radio emission
        (see Sect.~\ref{spixdisc} for details).
        }
\label{6cm}
\end{figure}

The map at 4.85\,GHz (Fig.~\ref{6cm}) shows primarily the bright jets of 4C\,70.19 in similar detail as the 145\,MHz LOFAR map. While no large-scale
emission is visible, the south-western extension from the end of the southern jet can be seen.
At the end of narrow and relatively straight parts of both jets knot-like brightenings (marked with arrows in Fig.~\ref{6cm}) are seen,
just before the emission becomes more extended. These features are similar to that observed in the radio galaxy NGC\,315 \citep{mack97} and will
be discussed later (see Sect.~\ref{morph}). Very interesting is the orientation
of the observed vectors of the magnetic field. Throughout the entire source it is perpendicular to the main axis, except
for a part of the northern jet, where it rapidly turns by around 90\degr and the vectors become parallel to the jet axis and follow
the direction of its propagation until the very end, where the vectors are again perpendicular to the jet axis, marking a possible termination
of the flow (see Sect.~\ref{propagation}).

The orientation of the B-vectors at 4.85\,GHz is, in fact, very different to what is seen in the map at 1.43\,GHz (Fig.~\ref{21cm}).
Because the relative rotation is close to 90\degr,
we have carefully checked if this could be related to instrumental effects or construction of the maps.
Therefore, we analysed both observations (VLA and Effelsberg) separately. We obtained the same results as from the combined data. Taking into account
that all polarised intensity and polarisation angle maps were produced in the same manner (also at other frequencies),
we conclude that the observed differences between 1.43 and 4.85\,GHz maps are of physical origin and should be attributed to Faraday rotation.
This will be further discussed in Sect.~\ref{magfields}.

\begin{figure}
        \resizebox{\hsize}{!}{\includegraphics[clip]{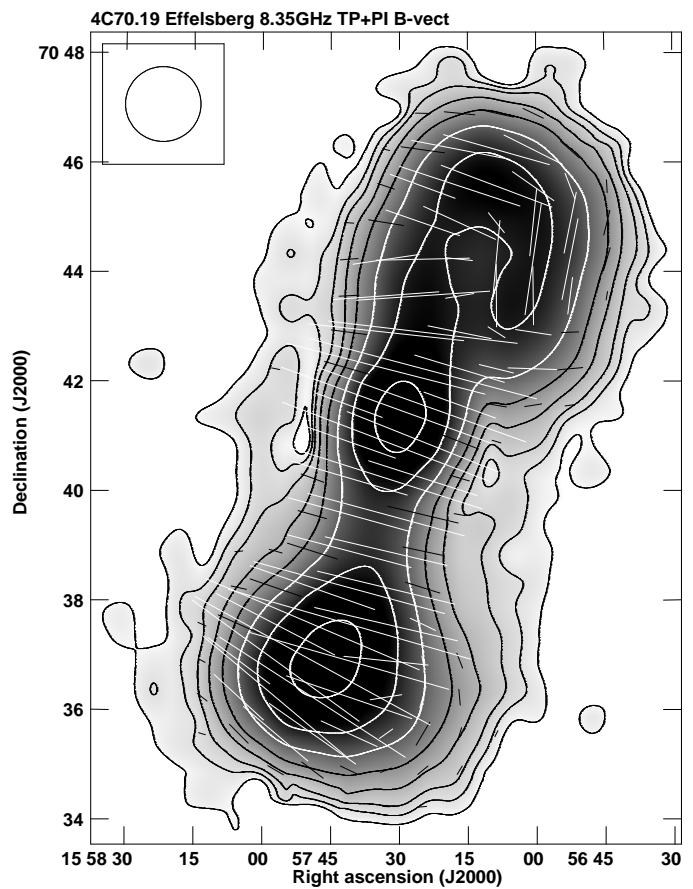}}
        \caption{Map of radio total power at 8.35\,GHz from the Effelsberg telescope. The contours are (3, 5, 8, 16, 32, 64, 128, and 256)\,$\times$\,0.31\,mJy\,beam$^{-1}$. The lines show the orientation of the magnetic field and a length of 1\arcmin correspond to polarised emission of 3.6\,mJy\,beam$^{-1}$.
        The beam size of 82\arcsec\ is presented in the upper left corner of the image.}
\label{4cm}
\end{figure}

The map at 8.35\,GHz (Fig.~\ref{4cm}), although having a relatively large beam ($\sim$\,1$\farcm$5), shows the same morphology as the map at 1.43\,GHz
(Fig.~\ref{21cm}). Furthermore, the sensitive single-dish observations allowed us to detect diffuse radio emission around the source,
visible not only at the positions of the radio lobes, but also around the central part of the source. Because of the large angular extent of
4C\,70.19 this cannot be attributed to beam-smearing of the brighter radio emission. The orientation of the observed B-vectors of the
magnetic fields is practically the same as that observed at 4.85\,GHz and despite the low resolution of the single-dish map,
its changes are also clearly visible.

\subsection{Large-scale diffuse radio emission}
\label{cocoon}

A comparison of all radio maps presented in this paper, as well as those available in the literature,
shows that the diffuse emission around 4C\,70.19 is only visible in the LOFAR map at 145\,MHz (Fig.~\ref{2m})
and in our single-dish Effelsberg map at 8.35\,GHz (Fig.~\ref{4cm}). Because the latter map shows emission at the highest radio frequency analysed
in this paper, these diffuse structures should also be well visible at lower frequencies.
Some signs of such emission are also visible in the maps at 327\,MHz, and 1.43\,GHz, but these maps do not show the large-scale structures.
We argue that this is entirely due to sensitivity of the maps. As we mentioned above the large-scale emission
outside of the radio jets in the 145\,MHz map is likely ``fractured'' by a small beam of these observations. Therefore,
we convolved this map to the beam of the 8.35\,GHz Effelsberg map. As expected, we obtained a map with practically the same extent of the large-scale
diffuse emission. This allowed us to estimate the spectral index of this large-scale emission and to calculate
the expected surface-brightness at each radio frequency studied in this paper.
The estimated sensitivities confirm that only in the single-dish Effelsberg
map at 8.35\,GHz the large-scale envelope around 4C\,70.19 can be detected at the 3$\sigma$ level.
We note here, that the sensitivities typically obtained with the Giant Metrewave Radio Telescope (GMRT) observations at 325 and 610\,MHz
\citep[e.g.][]{giacintucci11} should allow detection of the large-scale diffuse radio emission around 4C\,70.19. To date no such observations exist.

\subsection{Propagation of the jets}
\label{propagation}

\begin{figure}
        \resizebox{\hsize}{!}{\includegraphics[clip]{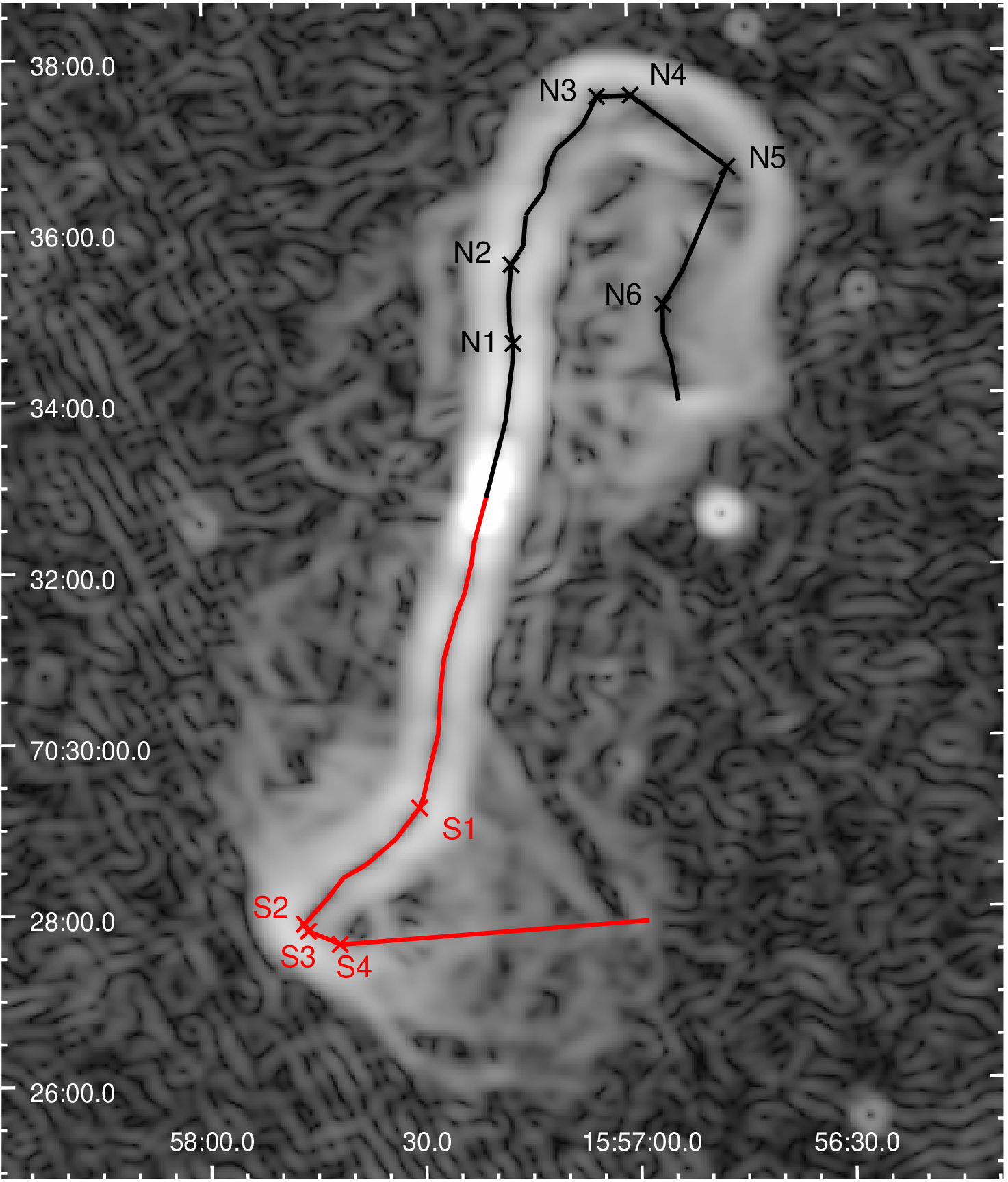}}
        \caption{LOFAR 145\,MHz map of 4C\,70.19 (Fig.~\ref{2m}) convolved to a beam of 15\arcsec\ and Sobel filtered with
        lines showing the propagation of the jets. The main turning points are marked with crosses and are labelled. See Sect.~\ref{propagation}.}
\label{profilemap}
\end{figure}

In Fig.~\ref{profilemap} we present the LOFAR map from Fig.~\ref{2m} but convolved to a resolution of 15\arcsec\ and processed using the {\it Sobel} filter
that enhances intensity gradients. This allowed to precisely follow the propagation of both jets of 4C\,70.19. This propagation is marked with black and red
lines for the northern and southern jet, respectively. We note here, that the direction of propagation of the southern jet beyond S4 turning point cannot
be clearly traced and the proposed line is likely one of the possible directions. Nevertheless, the proposed direction of the jet propagation seems
to be justified by the orientation of the B-vectors of the magnetic field observed in the high-frequency map of the polarised emission
(see the discussion in Sect.~\ref{magfields}).

\begin{figure}
        \resizebox{\hsize}{!}{\includegraphics[clip, angle=-90]{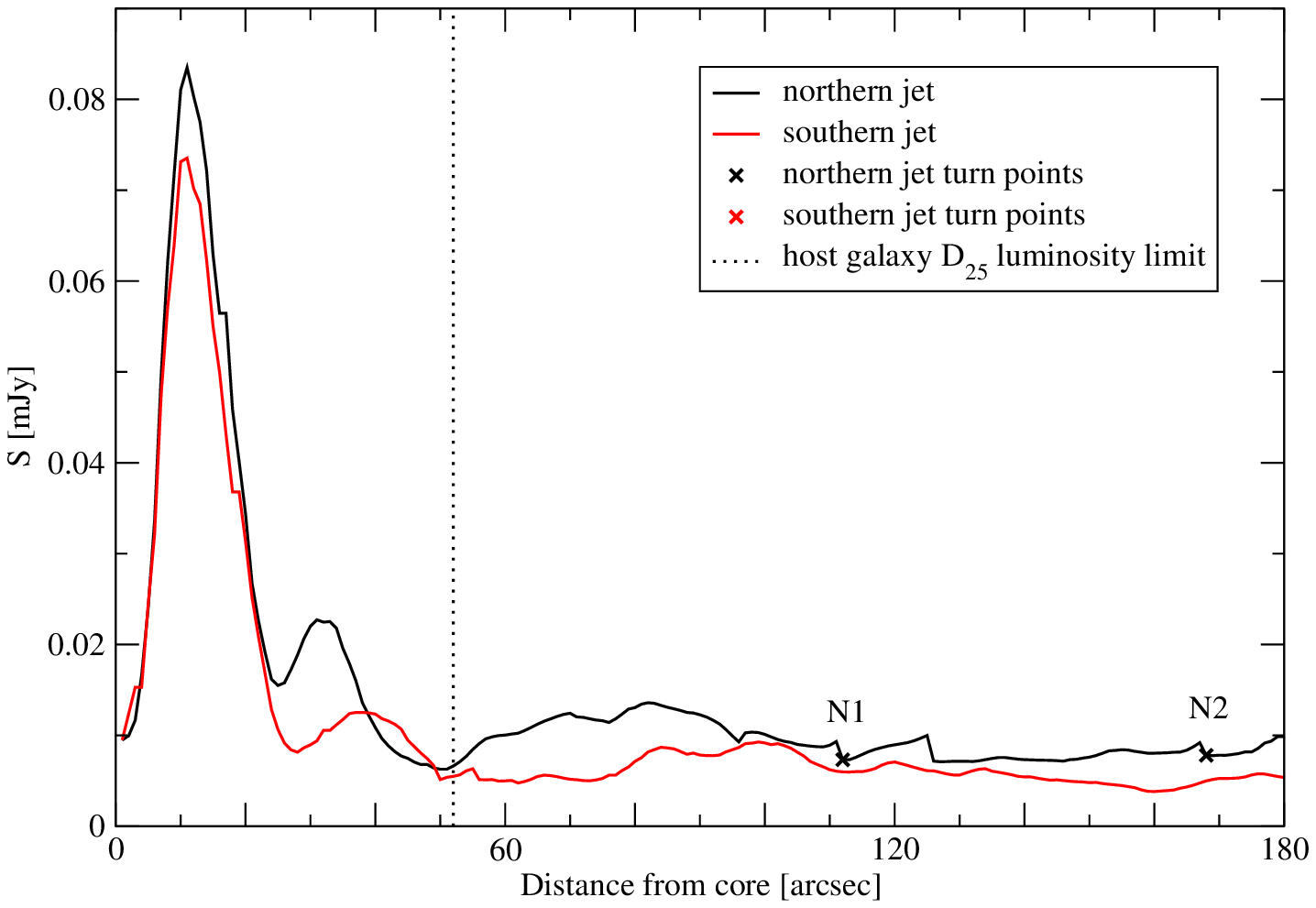}}
        \resizebox{\hsize}{!}{\includegraphics[clip, angle=-90]{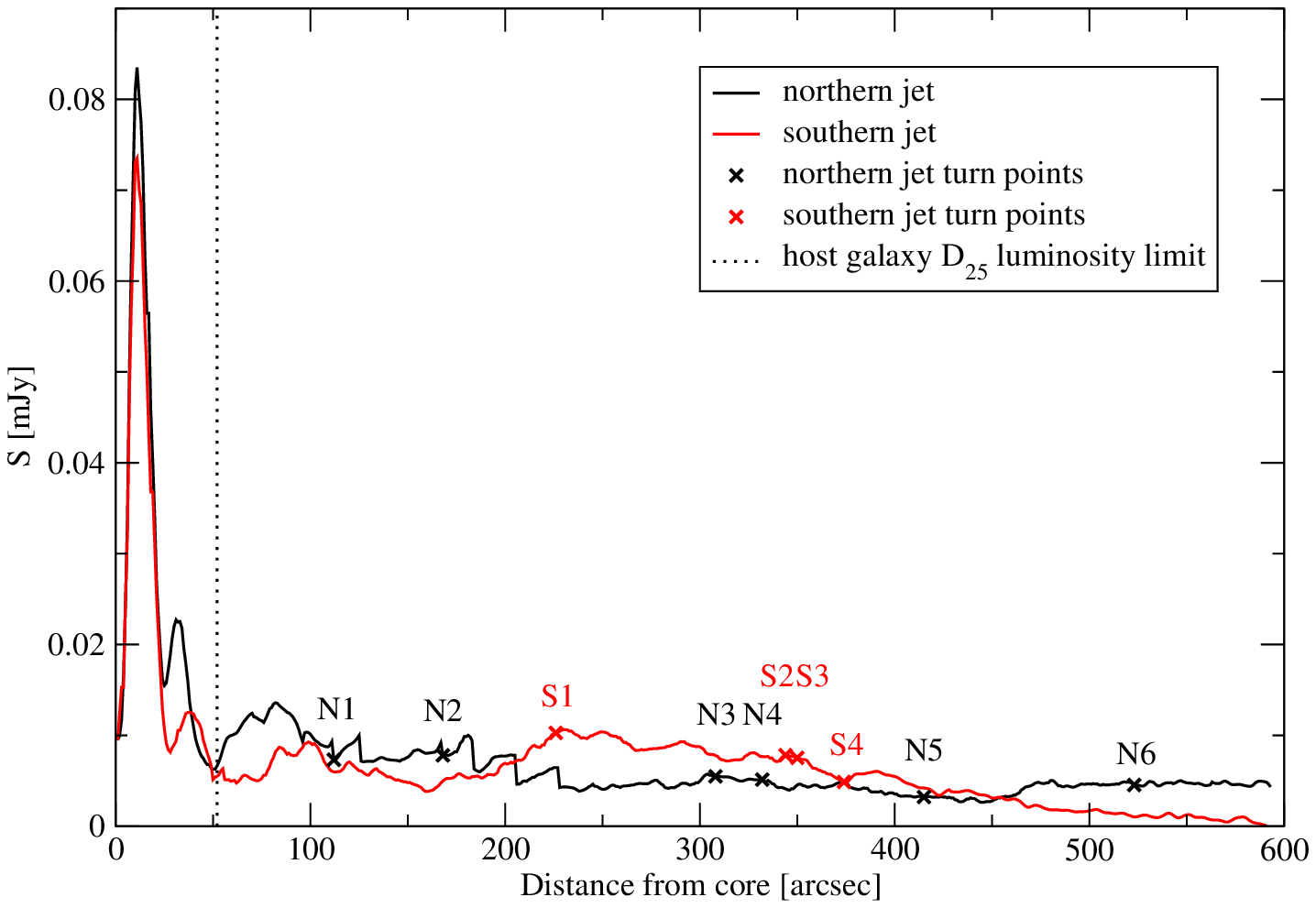}}
\caption{
	Intensity profiles of the jets of 4C\,70.19. Top: Intensity profiles measured in the LOFAR 145\,MHz map (Fig.~\ref{2m})
        and the combined VLA+Effelsberg maps at 4850\,MHz (Fig~\ref{6cm}) along the propagation lines presented in Fig.~\ref{profilemap}.
        The plot is limited to $3\arcmin$\,from the core to better present the inner parts of both jets.
        Bottom: Intensity profiles as in the upper panel, but along the entire propagation path.
        In both panels, the vertical dashed line marks the optical $D_{25}$ luminosity limit
        of the host galaxy NGC\,6048 and the main turning points of the jets
        are marked with crosses and are labelled in the corresponding colour.
        }
\label{intprofile}
\end{figure}

Frequent changes in the direction of the jet propagation can be noticed and the more significant ones are
marked with `X' and labelled. The profile of the intensity at 145\,MHz, measured along the propagation path, is presented in Fig.~\ref{intprofile}.
It shows distinct peaks around 10\,--\,15\arcsec\ from the centre, which correspond to 5\,--\,8\,kpc,
clearly confirming that the central source is not visible at this frequency and that instead two inner and highly symmetric radio features are detected,
being likely bases of the jets.
Two secondary maxima along both jets are visible around 30\,--\,40\arcsec\ (15\,--\,20\,kpc) with the northern one slightly closer to the core
and about two times brighter than the southern one.
The vertical dotted line marks the position of the ${\rm D}_{25}$ optical luminosity limit of the host galaxy.
Although this limit does not have a physical counterpart,
around this area a stronger gradient of brightness/ISM density of the host galaxy is expected. This finds its reflection in the profile of the
northern jet, whose intensity increases significantly.

The southern jet retains a relatively constant intensity along its path.
Around 200\arcsec\ (100\,kpc) from the core, both jets show the same (or comparable) intensities. Beyond this point the intensity
of the southern jet increases, until reaching the first turning point (S1), and of the northern jet decreases, roughly by the same amount.
Then, the intensity of the southern jet starts to steadily decrease and of the northern jet stays constant, regardless of the significant changes in the
direction of its propagation of both jets. It is also worth noting here that with the proposed propagation of the jets,
as inferred from the {\it Sobel}-filtered LOFAR map, both jets of 4C\,70.19 are in fact of similar length
(each reaching 300\,kpc, see lower panel of Fig.~\ref{intprofile}), and the asymmetric appearance results from projection effects.

\section{Discussion}
\label{disc}

\subsection{Galaxy environment}
\label{multi}

\begin{figure*}
	\sidecaption
        \includegraphics[width=12cm,clip]{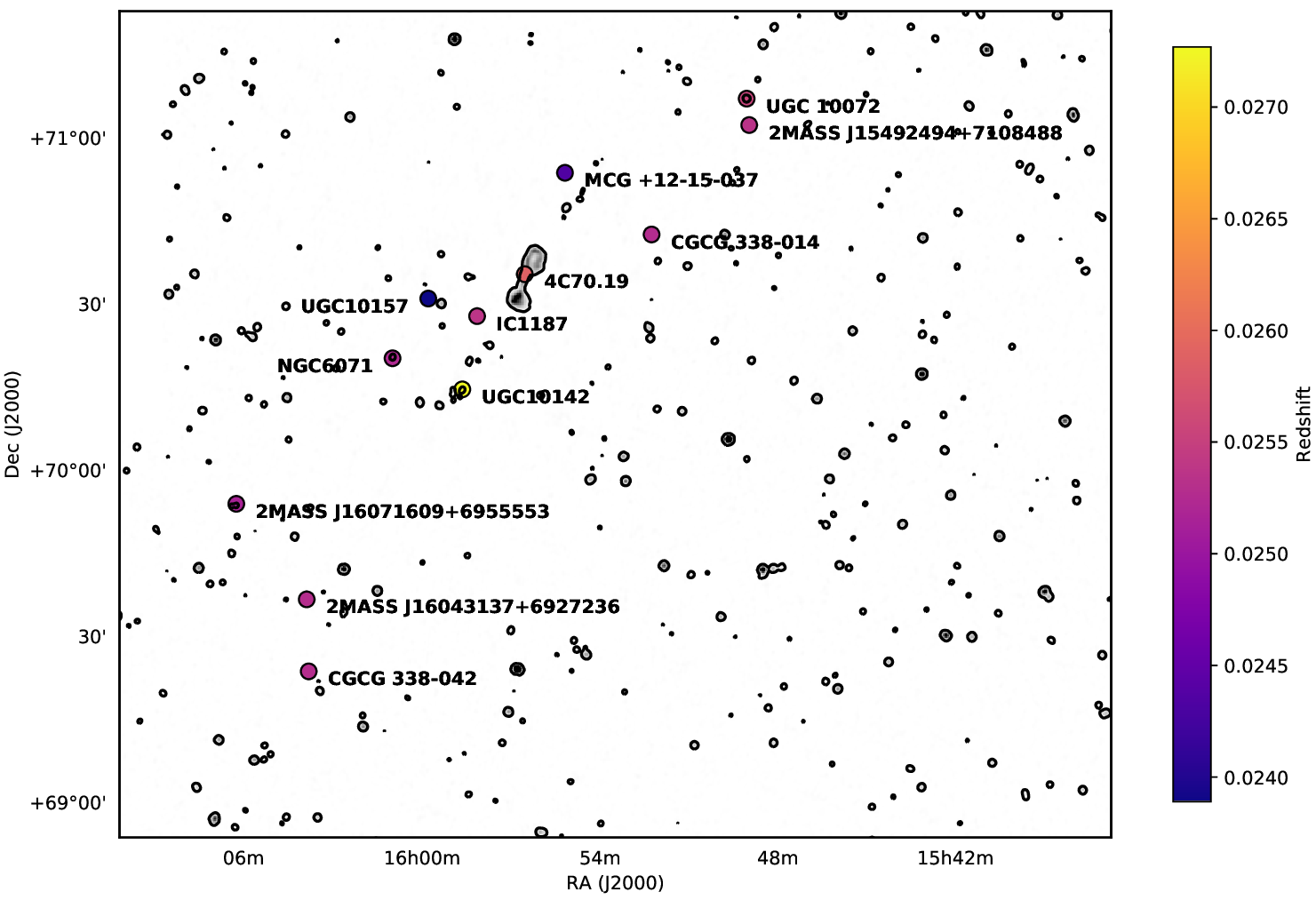}
        \caption{NVSS map of 4C\,70.19 at 1.4\,GHz with the group galaxies marked with filled circles and names. The redshifts of individual galaxies
        are colour coded according to the scale shown on the right.
        For a clear presentation of the large-scale radio map, only the contour at the level of 2.5\,mJy\,beam$^{-1}$ (5$\sigma$) is plotted.}
\label{group}
\end{figure*}

To investigate in more detail the origins of the distortions of 4C\,70.19,
we used three different catalogues of galaxy groups \citep{crook07,schombert15,tempel16}
and compiled a list of all possible members of the group that hosts NGC\,6048.
There are 12 spectroscopically confirmed candidate galaxies, where NGC\,6048 is the brightest group member.
Figure~\ref{group} shows the NVSS map (at 1.4\,GHz) of 4C\,70.19 with the group galaxies marked with circles and names.
The galaxy closest to NGC\,6048, at the distance of around 350\,kpc, is IC\,1187.

Although even this galaxy seems to be too distant for direct interactions with NGC\,6048,
there are two more objects, not shown in Fig.~\ref{group}, that might have influenced its evolution.

\begin{figure*}
\resizebox{0.56\hsize}{!}{\includegraphics[clip]{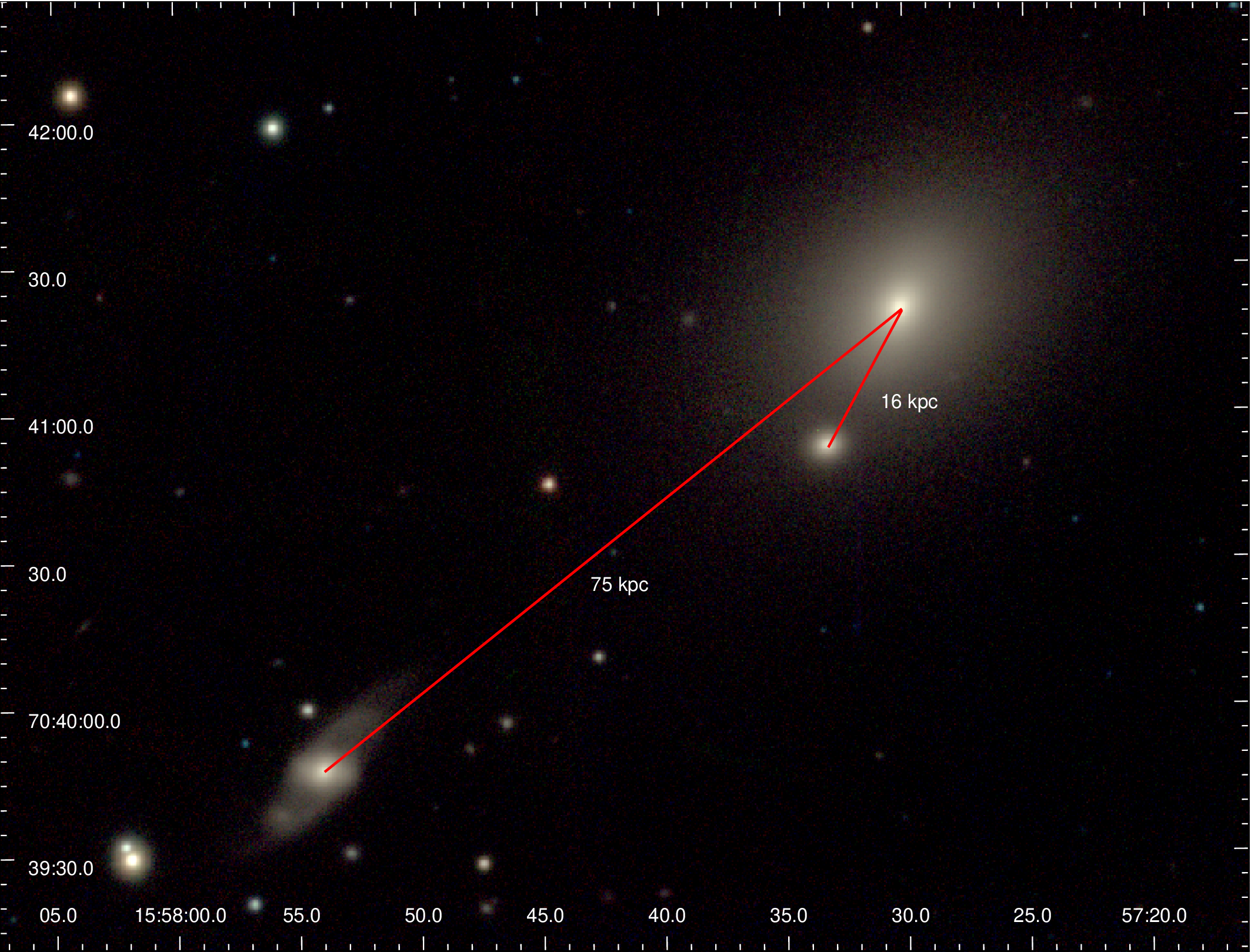}}
\resizebox{0.44\hsize}{!}{\includegraphics[clip]{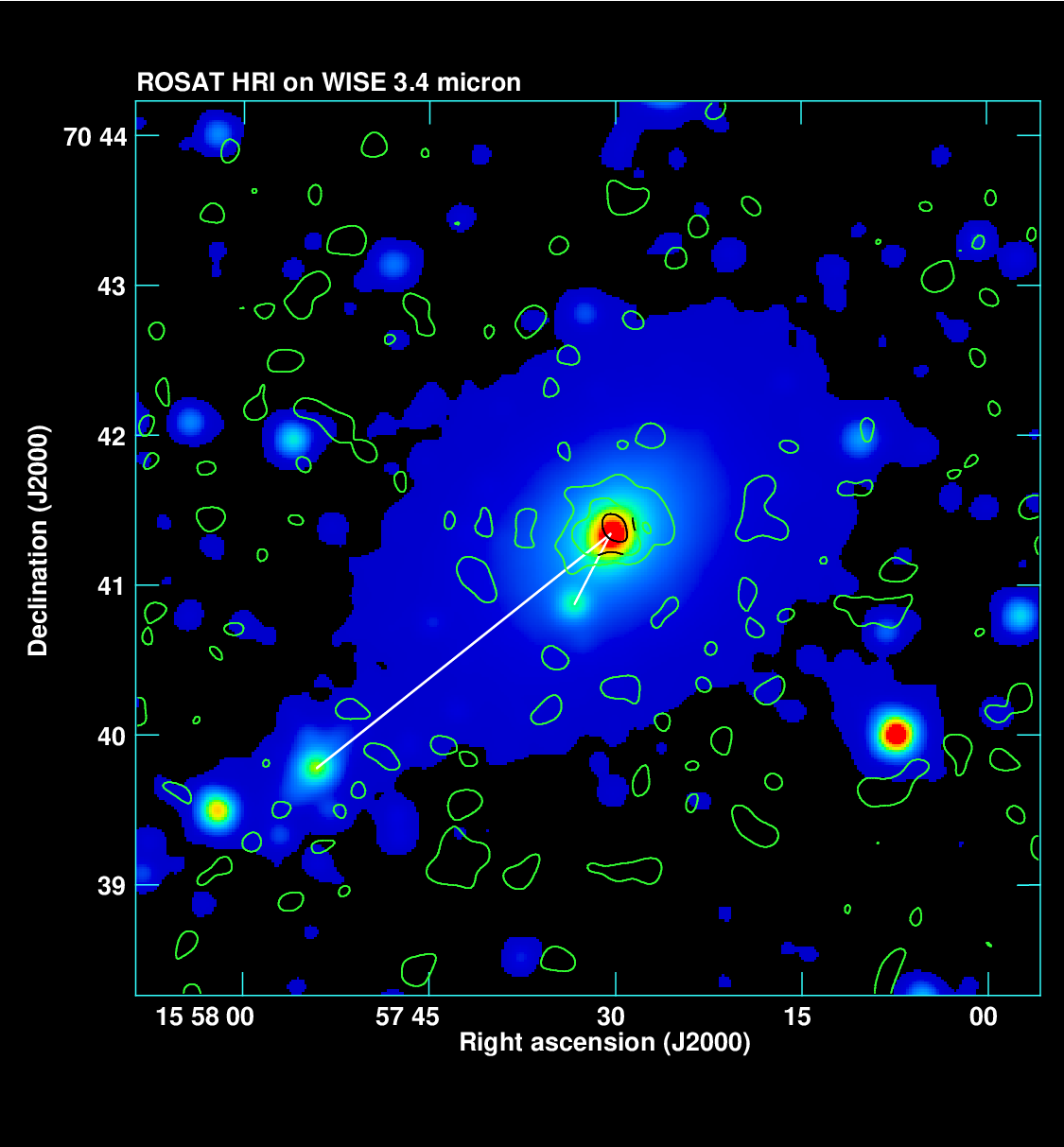}}
\caption{
        Optical images of NGC\,6048. Left: Pan-STARRS image of the NGC\,6048 system in the {\it r}, {\it i}, and {\it z} filters. 
	The red lines mark the projected distances
        between NGC\,6048 (top right) and the halo object, as well as the companion galaxy PGC214442 (bottom left).
        Right: WISE colour image at 3.4\,$\mu$m cut at 1$\sigma$ level overlaid with ROSAT HRI contours at 3, 5, and 8 $\times$ the r.m.s.. The white
        lines correspond to the red ones in the left panel.
}
\label{panstarrs}
\end{figure*}

The first of them, PGC214442 (RA=15$^{\rm h}$57$^{\rm m}$54$\fs$0, Dec=+70$\degr$39$^\prime$47$\farcs$14 (J2000.0)),
a galaxy of a total B-magnitude of 15.58$\pm$0.50 \citep{klemola87}, is visible southeast from NGC\,6048 (left panel of Fig.~\ref{panstarrs}).
It is described by \citet{devaucouleurs76}
to form a contact system with NGC\,6048. Unfortunately, for this galaxy no spectroscopic redshift has been measured.
Its photometric redshift was estimated to $z_{phot}=0.034\pm0.015$ by \citet{dalya18}. If both galaxies are located at the same distance,
their angular separation of $2\farcm5$ would correspond to a linear distance of only 75\,kpc. The optical images of PGC214442
from the Pan-STARRS data archive \citep{flewelling20} show the disturbed morphology of this galaxy (left panel of Fig.~\ref{panstarrs}).
It has a bright elliptical central part and a disc-like structure with a possible warp inclined to the major axis of the galaxy at around 70 degrees.
The disc is elongated towards NGC\,6048, which may result from interactions between these two galaxies.
Especially, that the halo around NGC\,6048, visible in the infrared image from WISE at 3.4\,$\mu$m (right panel of Fig.~\ref{panstarrs}), is also elongated
in the same direction. The infrared image is overlaid with contours of the ROSAT soft X-ray emission.
Using the ROSAT HRI countrate within $0\farcm5$ of the core region and assuming a simple power-law model
with $\Gamma$=1.6, we estimated the flux of the central source to be of the order of 3$\times$10$^{-13}$\,erg\,cm$^{-2}$\,s$^{-1}$.
At the distance to NGC\,6048 this results in a total luminosity of around 4.2$\times$10$^{42}$erg\,s$^{-1}$.
Of course, the countrate used includes all the emission from the hot gas in the disc of NGC\,6048; however, we do not expect
its contribution to be significant. Therefore, at few times 10$^{42}$erg\,s$^{-1}$ the central AGN of NGC\,6048 is of moderate luminosity,
confirming its low level of accretion (see Sect.~\ref{intro}). Both the dust halo around NGC\,6048 and the X-ray emission limited to
the optical extent of this giant elliptical (right panel of Fig.~\ref{panstarrs}) are likely a result of the recent merger (Sect.~\ref{intro}).

The second candidate source for the interactions is visible southeast from the core of NGC\,6048 (Fig.~\ref{panstarrs}).
Although no information about this compact object has been found in the literature, its brightness distribution in the image suggests
that this is an extra-galactic object rather than a foreground star and
so it might be physically present in the close vicinity of NGC\,6048. If the same distance to both sources is assumed,
they are separated by only 16\,kpc. However, for a clear verification of the nature and distance of this compact object,
spectroscopic observations are needed. Later in the paper we provide further arguments for the possible interaction of both sources with NGC\,6048.

\subsection{Radio morphology of the source}
\label{morph}

The distorted morphology of 4C\,70.19 resembles the radio galaxies IC\,708 \citep{vallee81} and 3C\,31 \citep{blandford78,heesen18}.
The winged shape of these two sources was explained as a result of projection effects arising from the orbital motion of the host galaxies
interacting with nearby massive objects.
As mentioned in Sect.~\ref{multi}, NGC\,6048 has two relatively close companion galaxies, which may influence its movement
through the IGM.
On the other hand, the 180\degr\,bend of one radio jet and an asymmetric broadening of the other makes this source even more similar to two other
radio sources, associated with elliptical galaxies NGC\,7626 and NGC\,315 \citep{giacintucci11}. These distortions of the radio lobes were explained
by the authors as interactions with the IGM. Particularly, the FRI source associated with NGC\,315 shows slight brightenings at the positions
where the radio plumes change their propagation direction \citep{giacintucci11}. Similar brightenings are visible (and marked with arrows)
in the map of 4C\,70.19 at 4.85\,GHz (Fig.~\ref{6cm}).

To find the inclination of the jet axis of 4C\,70.19 that could help to determine the orientation of the radio jets,
we analysed the orientation of both pairs of the inner features, using the intensity profiles and the formula used by e.g.
\citet{kuzmicz12}:

\begin{equation}
        i=\left[{\rm acos}\left(\frac{1}{\beta_j}\frac{(s-1)}{(s+1)}\right)\right].
\end{equation}

\noindent This formula assumes that the main cause of asymmetries in a radio source is Doppler boosting.
Here, $s=(S_j/S_{cj})^{1/2-\alpha}$ with $S_j$ and $S_{cj}$ are peak flux densities
of the structure closer and further from the core, respectively, and $\beta_j$ is the ratio of the jet velocity to the speed of light.
Following \citet{arshakian04} we assumed $\beta_j$ of 0.54 and the average spectral index $\alpha$ of
$-0.3$. To account for the ambiguous value of the core spectral index we also repeated our calculations with $\alpha=-0.5$.
Since the inner structures are almost perfectly symmetric,
we compared the calculations for each of them being the closer/further one. Because also the intensity of both features are similar, the
resulting inclinations varied by only around 8$\degr$\,with the major axis oriented almost exactly in the sky plane ($\sim$\,86\,--\,94$\degr$).

For the outer features, however, we obtained an inclination of 58$\degr$\,for $\alpha=-0.5$ and 65$\degr$\,for $\alpha=-0.3$.
Within our assumptions made for $\beta_j$ and $\alpha$, the 1\,mJy accuracy of the amplitudes in the profiles results in an uncertainty
of the derived angles of 1\,--\,2$\degr$. This means that the relative difference in orientation between the inner and the outer jet features
is of the order of 30$\degr$, with the northern jet of 4C\,70.19 being oriented towards us.
We note here that this change in orientation may be caused by a gas density gradient resulting from leaving the halo of the host galaxy,
as well as the existence of the halo object, southeast from the core (see Fig.~\ref{panstarrs}).

As a result, the outward orientation (with regard to the sky-plane) of the southern jet could produce an elongated hot spot-like feature
at the position of the jet bend, similar to that visible in the northern side. Then, the enhancement in the observed brightness
would be a result of integration of the emission from the bent jet.
This is in agreement with the intensity profiles along the jets (Sect.~\ref{propagation}),
which suggests that the south-eastern ``end'' of the southern jet is in fact its turning point.
This is supported by the increase in intensity of the central part of the southern lobe (lower panel of
Fig.~\ref{intprofile}), which leads to a higher flux integrated in this area and consequently a slightly higher value of the estimated magnetic field strength,
when compared to the corresponding part of the northern jet. If we assume that we see the entire northern jet in the sky plane,
the beginning of its bend (turning point N3 in Fig.~\ref{profilemap}) should be roughly
at the same distance from the core as the visible end of the southern jet (turning point S2 in Fig.~\ref{profilemap}).
The difference of both extents, as seen in the plot lies below 50\arcsec\ or 25\,kpc. We note here, however, that both jets deviate from the original axis
already closer to the core, as visible in Fig.~\ref{profilemap}. This makes more precise measurements of
full extent of the jets difficult and the difference below 25\,kpc could be attributed to projection effects, as well as differences in
the density of the surrounding medium that certainly influence the propagation of the jets.

\begin{figure*}
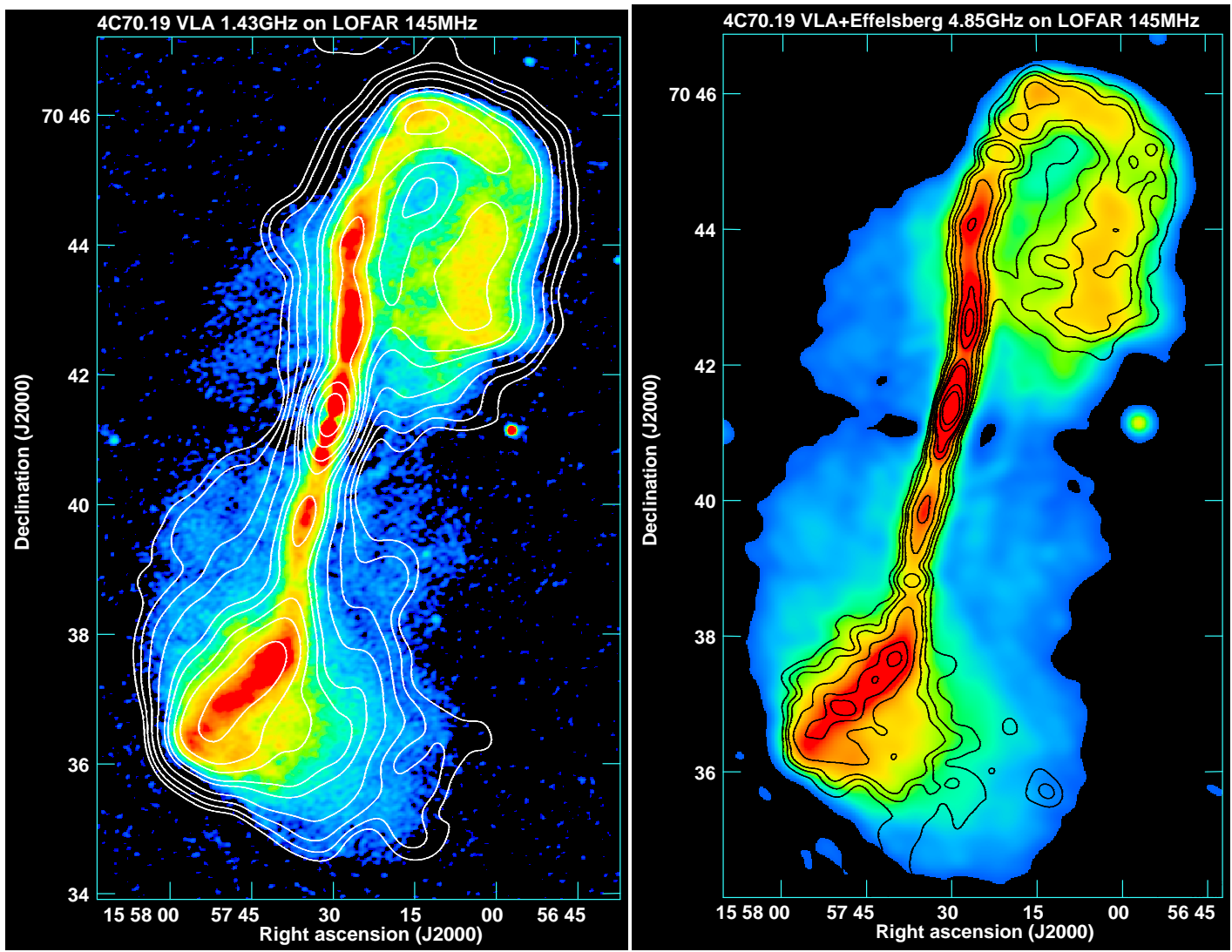

        \resizebox{0.51\hsize}{!}{\includegraphics[clip]{4c70.19_Lband_on_lofar6as_color.ps}}
        \resizebox{0.49\hsize}{!}{\includegraphics[clip]{4c70.19_Cband_on_lofar_color.ps}}
\caption{
        VLA maps of 4C\,70.19 overlaid on its LOFAR map. Left: VLA map of 4C\,70.19 at 1.43\,GHz from Fig.~\ref{21cm} 
	on the colour image of the LOFAR map at 145\,MHz from Fig.~\ref{2m}.
        The contour levels are (3, 6, 12, 24, 48, 64, 96, 128, 192, and 256)\,$\times$\,0.37\,mJy\,beam$^{-1}$.
        Right: VLA+Effelsberg map of 4C\,70.19 at 4.85\,GHz on the colour image of the LOFAR map at 145\,MHz.
        The latter map was convolved to 15\arcsec. The contours are (3, 6, 9, 16, 25, 32, 64, and 96)\,$\times$\,0.29\,mJy\,beam$^{-1}$.
        }
\label{vlalof}
\end{figure*}

Both panels of Fig.~\ref{vlalof} clearly show that the morphology of 4C\,70.19 practically does not change with frequency.
The very close correspondence of all radio features is obvious and both jets/lobes have the same extents in both maps at 145\,MHz and 4.85\,GHz, as
well as of the faint large-scale diffuse emission, whose parts are visible in the 1.43\,GHz map (left panel of Fig.~\ref{vlalof}).
This correspondence of the radio emission at both ends of the studied spectrum suggests that the observed edges of the emission
are rather of physical origin and are not caused by the observational limitations. A straightforward explanation would be the interaction with
the surrounding medium, whose pressure does not allow the jets to propagate further and with the combination of the orbital motion within the busy environment
of the galaxy group both jets bend. To some extent, this would resemble the famous radio source Virgo\,A, whose
jets interact with the surrounding intra-cluster medium \citep{degasperin12}. This in turn would suggest, that the faint large-scale emission visible
in the most sensitive maps (Sect.~\ref{cocoon}) is not a gaseous envelope or halo surrounding 4C\,70.19, but the plumes expanding behind the source.

\subsubsection{Magnetic fields}
\label{magfields}

\begin{figure}
        \resizebox{\hsize}{!}{\includegraphics[clip]{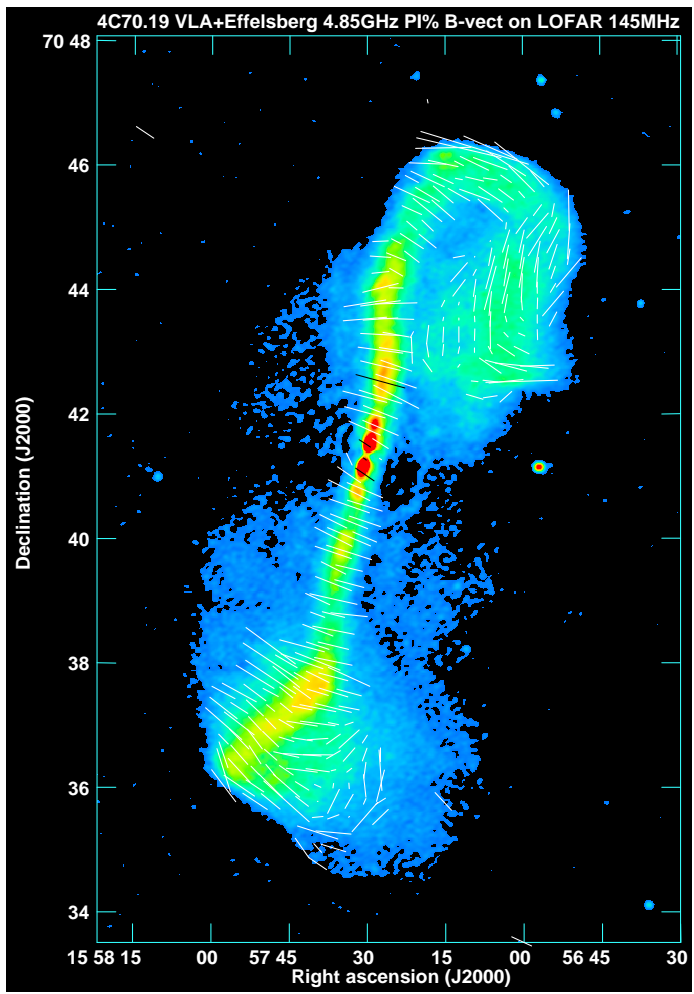}}
\caption{
        Map of polarisation degree in 4C\,70.19 at 4.85\,GHz produced from the Effelsberg single-dish map and the VLA interferometer maps in C and D
        configurations (Fig.~\ref{6cm}) overlaid on the LOFAR 145\,MHz map. The lines show the orientation of the magnetic field
        and a length of 1\arcmin corresponds to degree of polarisation of 50\%.
        }
\label{6deg}
\end{figure}

\begin{figure}
        \resizebox{\hsize}{!}{\includegraphics[clip]{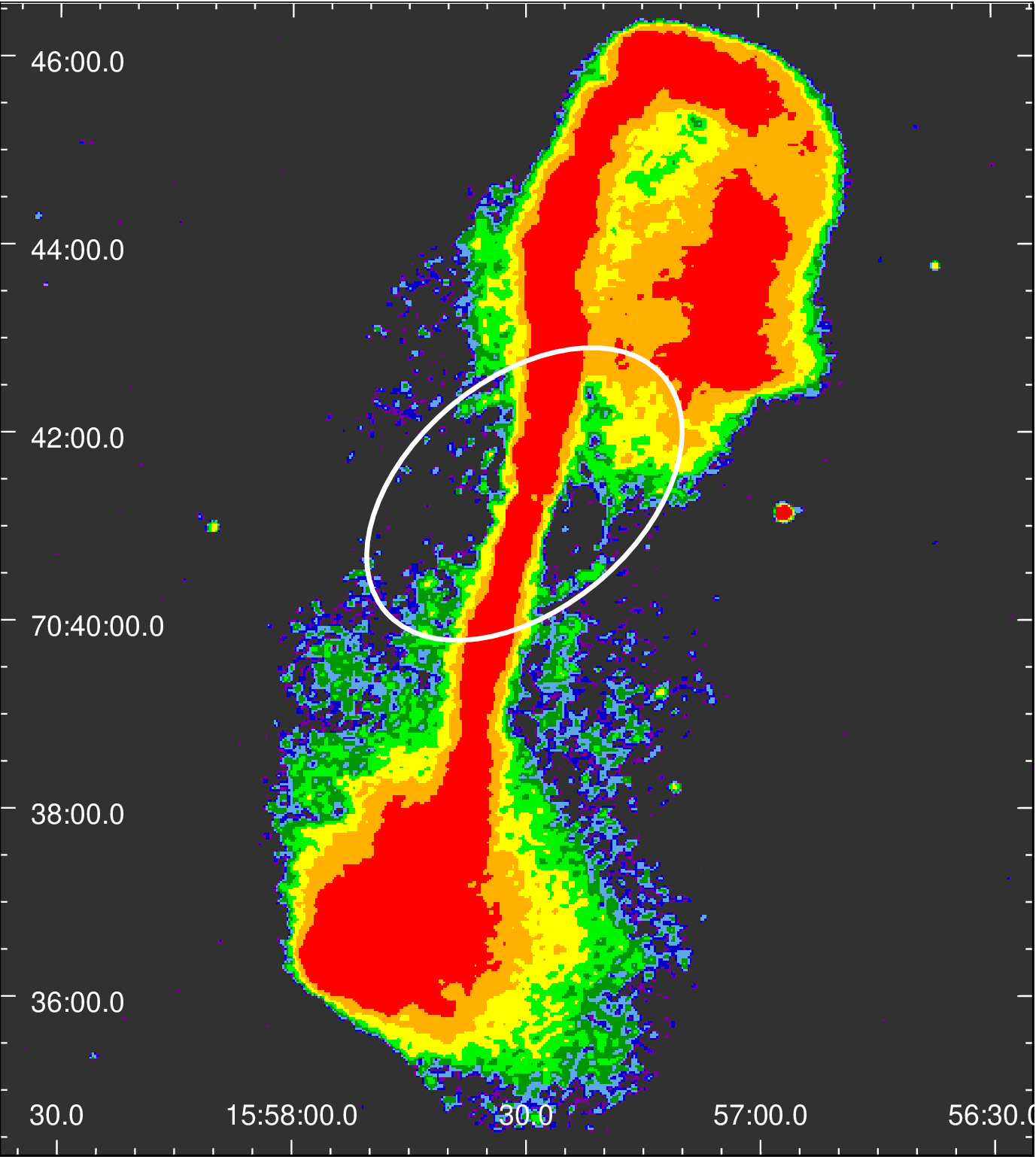}}
\caption{
        LOFAR map of 4C\,70.19 from Fig.~\ref{2m}. The white ellipse marks the extent of the infrared halo, as visible in the right panel of
        Fig.~\ref{panstarrs}.
        }
\label{dusthalo}
\end{figure}

As already shown in Sect.~\ref{radiomaps}, polarised radio emission obtained from our dedicated Effelsberg observations, as well
as the archival VLA data, allowed for a direct analysis of the orientation of magnetic fields in the sky plane. Especially interesting
is the behaviour of the magnetic field observed in the maps at high frequencies (Figs.~\ref{6cm} and~\ref{4cm}).
In the northern jet the orientation of the B-vectors is perpendicular to the jet axis until the bend
(points N3-N4 in Fig.~\ref{profilemap}). In the bend, the degree of polarisation in the outer
edge of the jet reaches almost 50\% (Fig.~\ref{6deg}) and the B-vectors become parallel to the jet axis. Towards the end of the jet, the B-vectors are
again perpendicular and the jet terminates with a notable brightening, well visible in the {\it Sobel}-filtered map (Fig.~\ref{profilemap}), and
an increase in the degree of polarisation (Fig.~\ref{6deg}).
Such changes of the orientation of the B-vectors are sensitive to gas motions and likely mark the area where the magnetic field
is being sheared (the bend) or compressed (jet termination). In such case, the bend of the jet would occur in the region, where the jet becomes too
weak to overcome the pressure of the surrounding medium and is eventually terminated after reaching even denser gas, while moving back
towards the host galaxy. The elongated halo around NGC\,6048 visible in the infrared map (right panel of Fig.~\ref{panstarrs})
has a shape of an ellipse, whose major axis is pointed almost exactly towards the companion galaxy PGC214442.
We overplotted this elliptical halo region on the LOFAR map of 4C\,70.19 in Fig.~\ref{dusthalo}. In the
north-west the halo reaches the end of the bent northern jet, which could explain the addtional kink and its end, as well as the termination shock,
or the deceleration point, beyond which the plume expands freely behind the source in the scenario proposed in the previous section.

In the southern jet, the B-vectors of the magnetic field are perpendicular to its axis along the entire line of propagation, also after the slight bend
marked in Fig.~\ref{profilemap} as S1. Nevertheless, if we follow the propagation line beyond the last turning point marked (S4), the B-vectors
become parallel to the (proposed) jet axis, just like in the northern jet. Also in this case we would expect shearing of the magnetic
field at the bend of the jet. This is supported by a slight increase in the degree of polarisation around turning points S2 and S3 (Fig.~\ref{6deg}).
These findings provide additional arguments for the assumed propagation of the southern jet and that the extension to the south-west is
not a diffuse radio lobe but rather the southern jet visible from the side after the bend and extending into a radio plume.

Because we excluded the possibility that the roughly 90\degr\ difference in the orientation of the B-vectors seen
between 1.43\,GHz (Fig.~\ref{21cm}) and 4.85\,GHz (right panel of Fig.~\ref{6cm}) is of instrumental
origin (Sect.~\ref{radiomaps}), this difference needs a physical explanation.
Especially, that the global orientation of the magnetic field visible
in the low-resolution map at 2.67\,GHz (Fig.~\ref{11cm}) seems to be intermediate between those at 1.43 and 4.85\,GHz. Both the LOFAR map at
145\,MHz (Fig.~\ref{2m}) and the sensitive single-dish Effelsberg map at 8.35\,GHz (Fig.~\ref{4cm}), as well as our analysis of the surroundings of the host
galaxy NGC\,6048, presented in the previous section, suggest the presence of magnetised plasma, which could produce Faraday rotation
responsible for the observed changes in the orientation of the polarisation angle.
The polarised radio data at 4.85 and 8.35\,GHz, where the expected Faraday depolarisation is low,
allowed us to obtain a map of RM distribution (Fig.~\ref{rm}).
The RM values are roughly uniform throughout the source with a mean of -35\,rad\,m$^{-2}$. The only exception is
the area where the northern lobe starts to bend. The rotation measure is around 0\,rad\,m$^{-2}$ at this position.
RMs of the order of few tens of\,rad\,m$^{-2}$ do not significantly affect the polarisation angle
at high frequencies, resulting in uncertainties of only few degrees at 4.85 or 8.35\,GHz. 
{To estimate the relative rotation of the polarisation plane between these high frequencies and 1.43\,GHz, 
we use the simple relation between the rotation angle, rotation measure, and wavelength ($\chi = {\rm RM}\times\,\lambda^2$). Then, 
at 1.43\,GHz a rotation measure of (-)35\,rad\,m$^{-2}$ results in a rotation 
of the polarisation plane by around 88\degr, which easily explains the differences of the orientation 
of the magnetic field vectors mentioned before.}

The observed RMs result from the superposition of the intrinsic Faraday rotation around 4C\,70.19 and
all 'Faraday screens' along the line-of-sight towards the source. We used the all-sky catalogue of RMs \citep{taylor09} to estimate the contribution
only from 4C\,70.19. The average RM within 5\degr\ from the source is +7.6$\pm$2.8\,rad\,m$^{-2}$, which means that the RM intrinsic to the source
are between --\,45 and --\,40\,rad\,m$^{-2}$. Using the strengths of the magnetic field in the radio source, these can be used to estimate
local electron densities with the use of the well-known formula for the Faraday-thin source:

\begin{figure}
        \resizebox{\hsize}{!}{\includegraphics[clip]{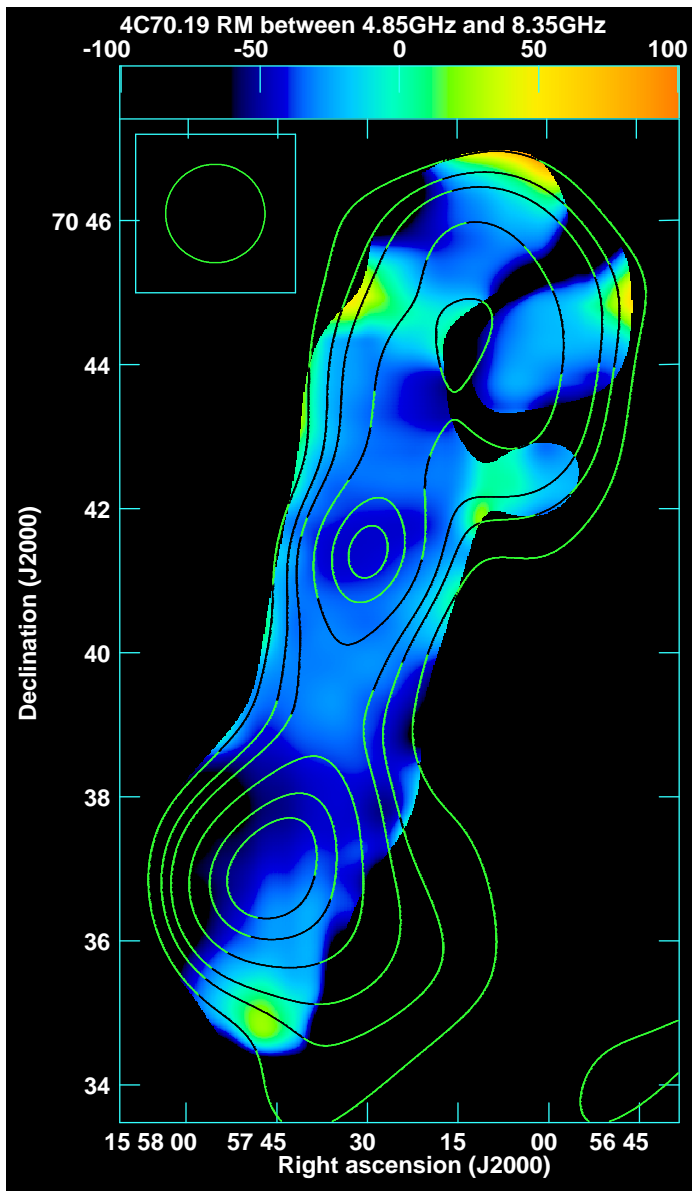}}
\caption{
        Rotation measure map of 4C\,70.19 between 4.85 and 8.35\,GHz. The map is overlaid with contours of the total radio emission map
        at 4.85\,GHz (Fig.~\ref{6cm}) convolved to the resolution of the 8.35\,GHz map from Fig.~\ref{4cm}.
        The contour levels are (3, 6, 9, 16, 25, 32, 64, and 96)\,$\times$\,3.57\,mJy\,beam$^{-1}$.
        The beam size of 82\arcsec\ is presented in the upper left corner of the image.
        }
\label{rm}
\end{figure}

\begin{displaymath}
        {\rm RM}[{\rm rad}\,\,{\rm m}^{-2}] = 812\int^{L}_{0}n_e[{\rm cm}^{-3}]{B}[\mu\,{\rm G}]\cdot d{l}[{\rm kpc}],
\end{displaymath}

\noindent {where RM is rotation measure, $n_e$ is the electron density of the medium, 
${B}$ is the magnetic field component along the line of sight, and $d{l}$ is the pathlength through the medium.}

The strengths of the magnetic field were calculated with the assumption of the minimum energy conditions and following \citet{longair11}.
Our assumptions included the cutoff frequencies of $v_{min}$=10\,MHz and $v_{max}$=100\,GHz,
the filling factor of 1, and the pure electron-positron plasma. The volumes
were approximated by a cylindrical shape of 164\arcsec\ (twice the beam size) in length and a radius of 41\arcsec.
For the core region we used a spherical shape with a diameter of 82\arcsec.
Local values of the magnetic field strength in different regions of the source are presented in Fig.~\ref{spixchart}.
The typical magnetic field strengths are 1.5\,$\mu$G in most regions of the source.
The only exception is the core region (2.1\,$\mu$G) and the central region of the southern lobe (bottom middle graph in Fig.~\ref{spixchart}, 1.8\,$\mu$G).
We note here, however, that the uncertainty of the derived strengths of the magnetic field is about 0.5\,$\mu$G.

We used the obtained value of the magnetic field strength to estimate the electron densities of the halo/IGM gas around 4C\,70.19 required to
produce the RMs observed in our maps. With the above formula, RMs around 40\,rad\,m$^{-2}$
produced within 4C\,70.19 (we used the same volumes as for the magnetic field estimates), require electron densities
of around 10$^{-3}$\,cm$^{-3}$. Such densities fit well between low-density interstellar medium
of galaxies (few 10$^{-3}$\,cm$^{-3}$) and the IGM
(a few 10$^{-4}$\,cm$^{-3}$). The derived densities are in fact only lower limits, because our estimates of the magnetic field strengths
correspond to the brightest parts of the radio emission, close to the jets.
Furthermore, only the line of sight component of the magnetic field is responsible for the observed RMs. Taking both into account,
lower values of the magnetic field strengths would require a higher density of the medium to produce the observed RMs. This would make the density of the
IGM around 4C\,70.19 even closer to the typical values of the low-density ISM, providing another argument for the interactions
of NGC\,6048 with neighbouring galaxies, as already sugested by the extended and elongated infrared halo (right panel of Fig.~\ref{panstarrs} and
Fig.~\ref{dusthalo}).

\subsubsection{Spectral properties}
\label{spixdisc}

\begin{figure}
        \resizebox{\hsize}{!}{\includegraphics[clip, angle=-90]{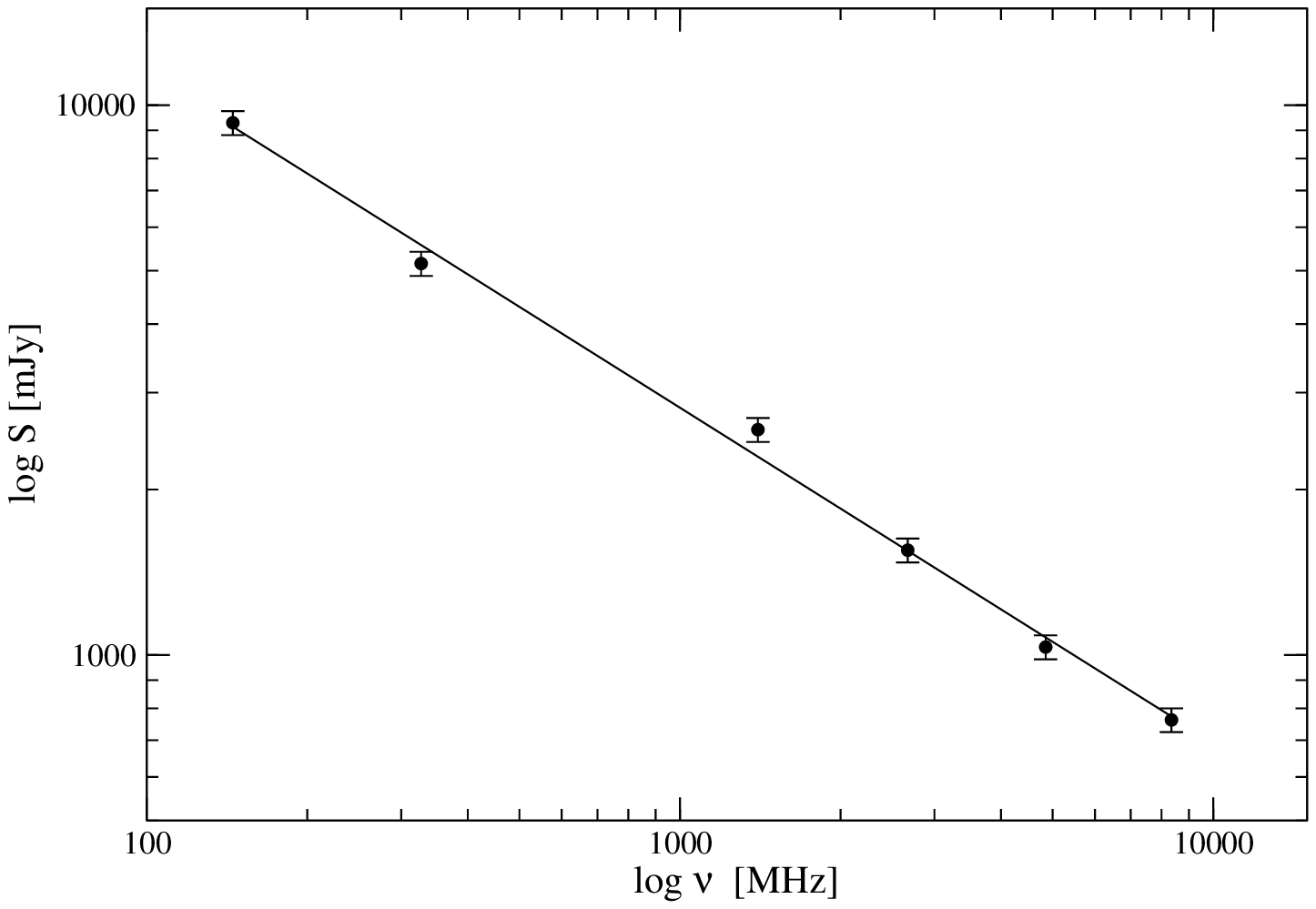}}
        \caption{Integrated spectrum of 4C\,70.19.}
\label{globalspix}
\end{figure}

We used all radio data presented in this paper to construct a global radio spectrum of the source (Fig.~\ref{globalspix}).
The mean spectral index $\alpha$ in the 145\,MHz -- 8350\,MHz range is $-0.61\pm0.02$ ($\rm S\propto \nu^{\alpha}$).
We note here that in the two radio maps observed with the single-dish Effelsberg telescope, that is, at 2.67 and
4.85\,GHz, the radio morphology of the source was not clearly visible due to a large beam of the observations. This
made precise flux measurements difficult because of possible contamination by the background sources.
Nevertheless, the map at 4.85\,GHz was used only for combination with the high-resolution VLA map (see
Sects.~\ref{archivalradio} and~\ref{radiomaps}), hence no direct flux measurement was performed.
For the map at 2.67\,GHz with a beam as large as $4\farcm4$, we decided to measure the flux only in the area where
the emission in the LOFAR map at 145\,MHz convolved to the resolution of the Effelsberg map at
8.35\,GHz is above 3$\sigma$. The obtained total flux at 2.67\,GHz perfectly fits the global spectrum of 4C\,70.19 (Fig.~\ref{globalspix}).

\begin{figure*}
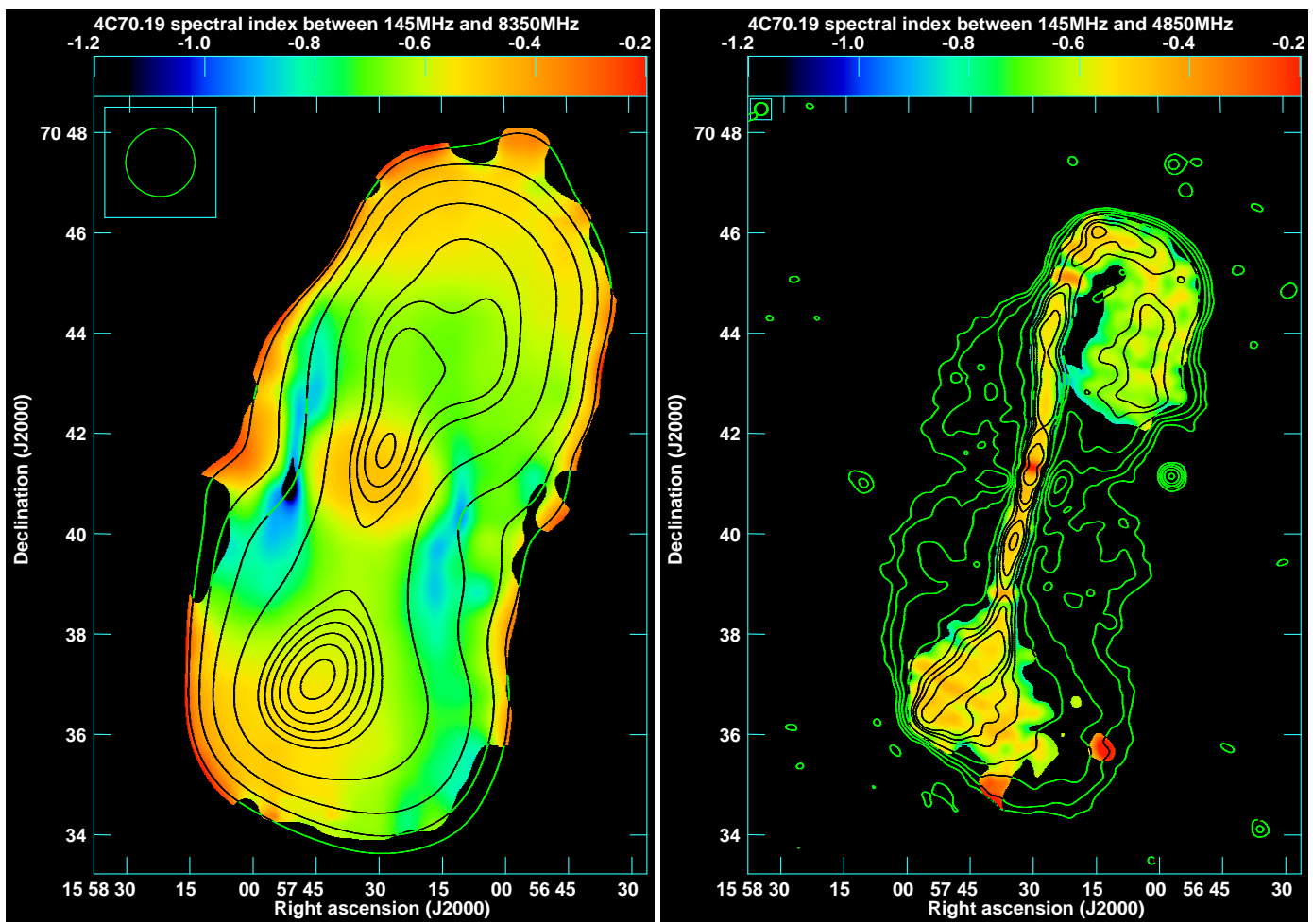

        \resizebox{0.49\hsize}{!}{\includegraphics[clip]{4c70.19_spix_lowres.ps}}
        \resizebox{0.49\hsize}{!}{\includegraphics[clip]{4c70.19_spix_highres.ps}}
\caption{
	Spectral index maps of 4C\,70.19. Left: Map between 145\,MHz (LOFAR) and 8350\,MHz (Effelsberg) overlaid with contours of the former map.
        The contour levels are (3, 10, 30, 100, 250, 300, 350, 400, 450, and 500)\,$\times$\,1.8\,mJy\,beam$^{-1}$.
        The beam size of 82\arcsec\ is presented in the upper left corner of the image.
	Right: Map between 145\,MHz (LOFAR) and 4850\,MHz (VLA+Effelsberg) overlaid with contours of the LOFAR map
        convolved to the resolution of the latter map. The contour levels are (3, 8, 16, 32, 64, 80, 96, 128, and 256)\,$\times$\,0.32\,mJy\,beam$^{-1}$.
        The beam size of 15\arcsec\ is presented in the upper left corner of the image.
        }
\label{spix}
\end{figure*}

To investigate the distribution of the spectral index throughout the source,
we constructed two spectral index maps using the LOFAR 145\,MHz, the combined Effelsberg+VLA map at 4.85\,GHz,
and the Effelsberg 8.35\,GHz radio maps. This allowed to obtain a high resolution view of the spectral index in jets and lobes,
as well as the large-scale diffuse emission (Fig.~\ref{spix}).
Usually, a relatively flat global spectral index (as calculated from the total fluxes measured in our maps)
suggests a flat spectrum of the central region and the lobes/hot spots
accompanied by a very steep spectrum of the radio cocoon \citep{miley80}.
In the low resolution spectral index map (left panel of Fig.~\ref{spix}), which includes the large-scale diffuse emission,
the areas of the steeper spectral index are found not only close to the core region, {suggesting backflow,
\citep[e.g. similarly as in the case of 3C\,296 and B2\,0326+39,][]{laing11}, but also in the south-western outskirts 
of the lobe \citep[e.g. similarly as in the case of the northern extension of NGC\,6251,][]{cantwell20}.} This could mean that the backflow
was acting not only towards the centre but also perpendicular to the jet axis, also pointing at the possible
bending of the southern jet. Another possibility is that this steepening of the spectral index in the
south-west marks the outer parts of the jet, where the emission extends in a plume.

We also note here that the areas of the very flat or even inverse spectral index at the northern and southeastern boundaries of 4C\,70.19
(left panel of Fig.~\ref{spix}) suggest that some of the faint diffuse emission may be lost in the interferometric LOFAR observations.
This can be also seen as steep gradients in the emission distribution in the LOFAR map (Fig.~\ref{2m}).
Nevertheless, such gradients are visible, within the resolution limits, in all our radio maps, especially in the south-eastern part,
which could confirm the bending of the southern jet, so that the visible gradient is just a result of a tangential motion to line of sight.

\begin{figure}
        \resizebox{\hsize}{!}{\includegraphics[clip,angle=-90]{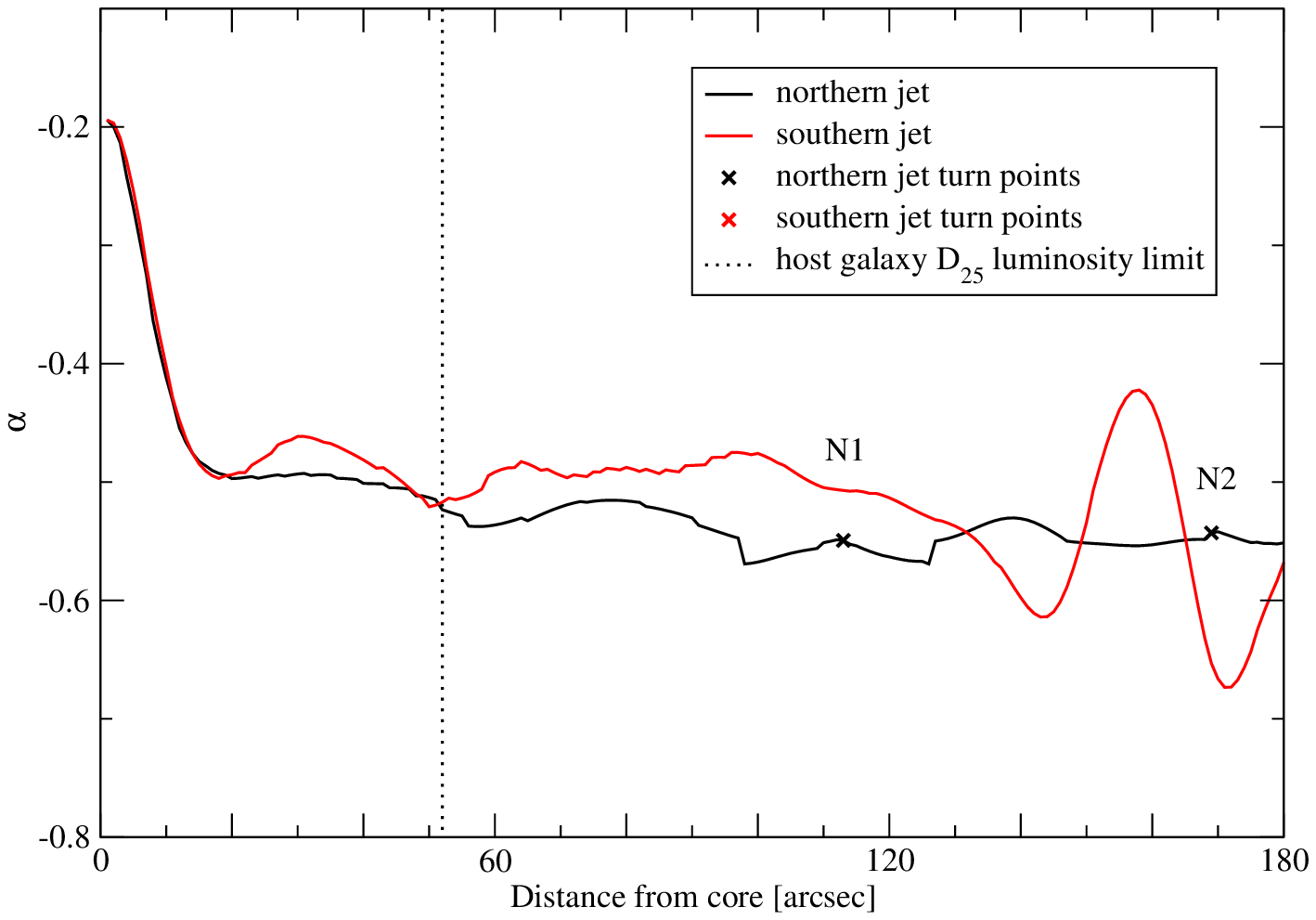}}
        \resizebox{\hsize}{!}{\includegraphics[clip,angle=-90]{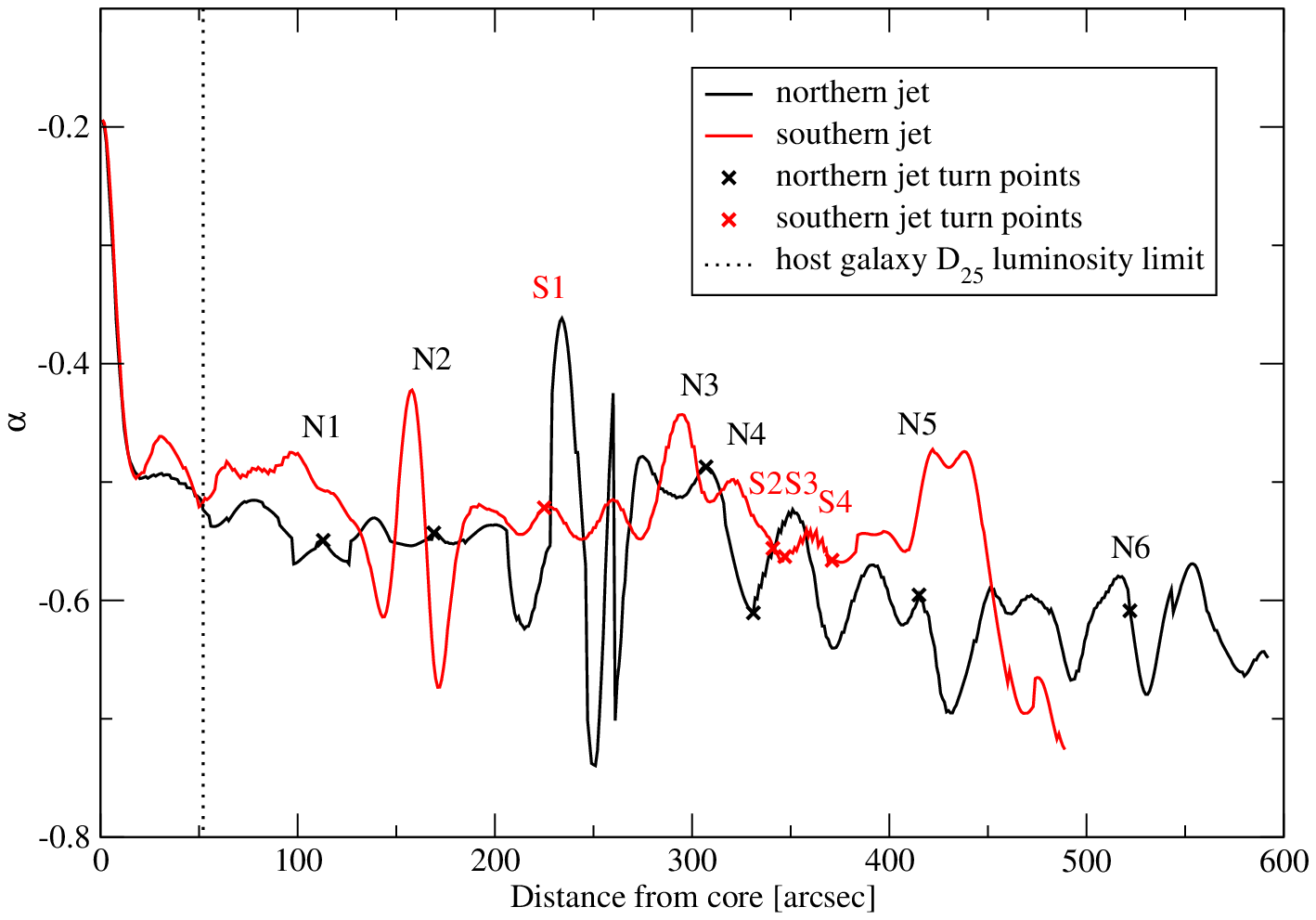}}
\caption{
        Spectral index profiles along the jet propagation lines of 4C\,70.19. Top: Profiles measured in the spectral index map between 
	145 and 4850\,MHz (left panel of Fig.~\ref{spix}).
        The plot is limited to $3\arcmin$\,from the core to better present the inner parts of the jets.
        Bottom: Profiles as in the upper panel, but along the entire propagation path.
        In both panels the vertical dashed line marks the optical $D_{25}$ luminosity limit
        of the host galaxy NGC\,6048 and the main turning points of the jets
        are marked with crosses and labelled in corresponding colour.
        }
\label{spixprofile}
\end{figure}

A more detailed view of the spectral index along the jets of 4C\,70.19 is provided by the spectral index map between the LOFAR 145\,MHz
and the combined 4.85\,GHz maps (left panel of Fig.~\ref{spix}) with a resolution of 15\arcsec. While the very flat spectral index
of the core is not surprising, because it is not visible in the low-frequency map, very interesting are the two compact areas of a flat spectral
index found at the end of the (relatively) straight parts of both jets, just before they turn and extend into more diffuse structures.
We constructed a profile of the spectral index similar to that of the total intensity (Fig.~\ref{spixprofile}). Although no significant changes
of the spectral index with regard to the turning points can be observed, it is worth to note here that both compact areas of a flat spectral index mentioned
above are found in almost the same linear distances from their corresponding turning points (N3 and S1 in Fig.~\ref{profilemap}).
These compact regions mark the enhancements of the radio emission visible in the map at 4.85\,GHz as bright knots of emission,
shown with arrows in Fig.~\ref{6cm}.
It is possible that in these two regions the jets encounter locally denser medium and decelerate, which causes their subsequent turns and
diffusion into plumes.

As we already mentioned in Sect.~\ref{morph}, similar bright knots of radio emission,
beyond which the jets start to extend into diffuse plumes were also observed
in the radio galaxy NGC\,315 \citep{mack97}. The host of this radio galaxy also resides within a group of galaxies and therefore the presence
of IGM with varying density is very probable.

We compared the spectral index of 4C\,70.19 with those of the radio galaxies studied by \citet{giacintucci11}, who investigated interactions of
AGNs with the IGM of their host galaxy groups. Six (out of 15) sources from this sample show spectral indices between
235 and 610\,MHz of around -0.8 or flatter. Four of them have small linear sizes
and are limited to the vicinity of their host galaxies, which suggests relatively young age of these sources.
The remaining sources, NGC\,7626 and 3C\,31, each with a spectral index of -0.81, have larger linear sizes of 185 and 900\,kpc,
respectively \citep{giacintucci11}. 4C\,70.19, having a global spectral index of -0.61 and the linear size of 330\,kpc, can be
therefore considered a similar object. Nevertheless, while 3C\,31 has a radio power practically identical to 4C\,70.19,
NGC\,7626 is one order of magnitude weaker. Although both 3C\,31 and NGC\,7626 show distorted radio tails,
their morphology seems to be less complex than that of 4C\,70.19 and the bends of their jets are likely observed roughly in the sky-plane.
In fact, if we assume the real total extent of 4C\,70.19 to be around 600\,kpc, as our analyses presented in previous sections suggest,
it corresponds to sources like 3C\,31 even more closely.

\begin{figure}
        \resizebox{\hsize}{!}{\includegraphics[clip]{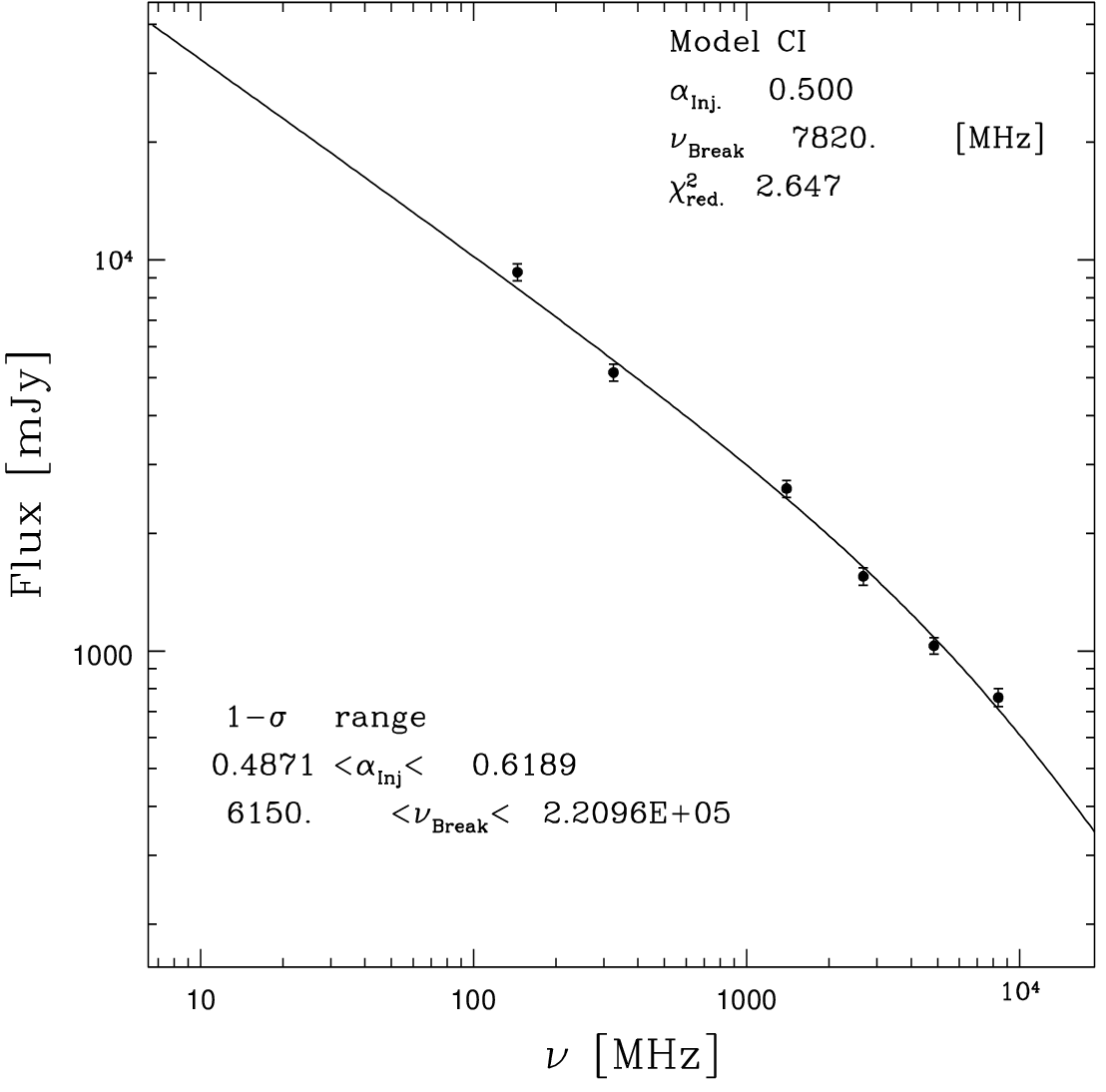}}
        \caption{Modelled global spectrum of 4C\,70.19.}
\label{globalspixCI}
\end{figure}

\begin{figure*}
\resizebox{\hsize}{!}{\includegraphics[clip]{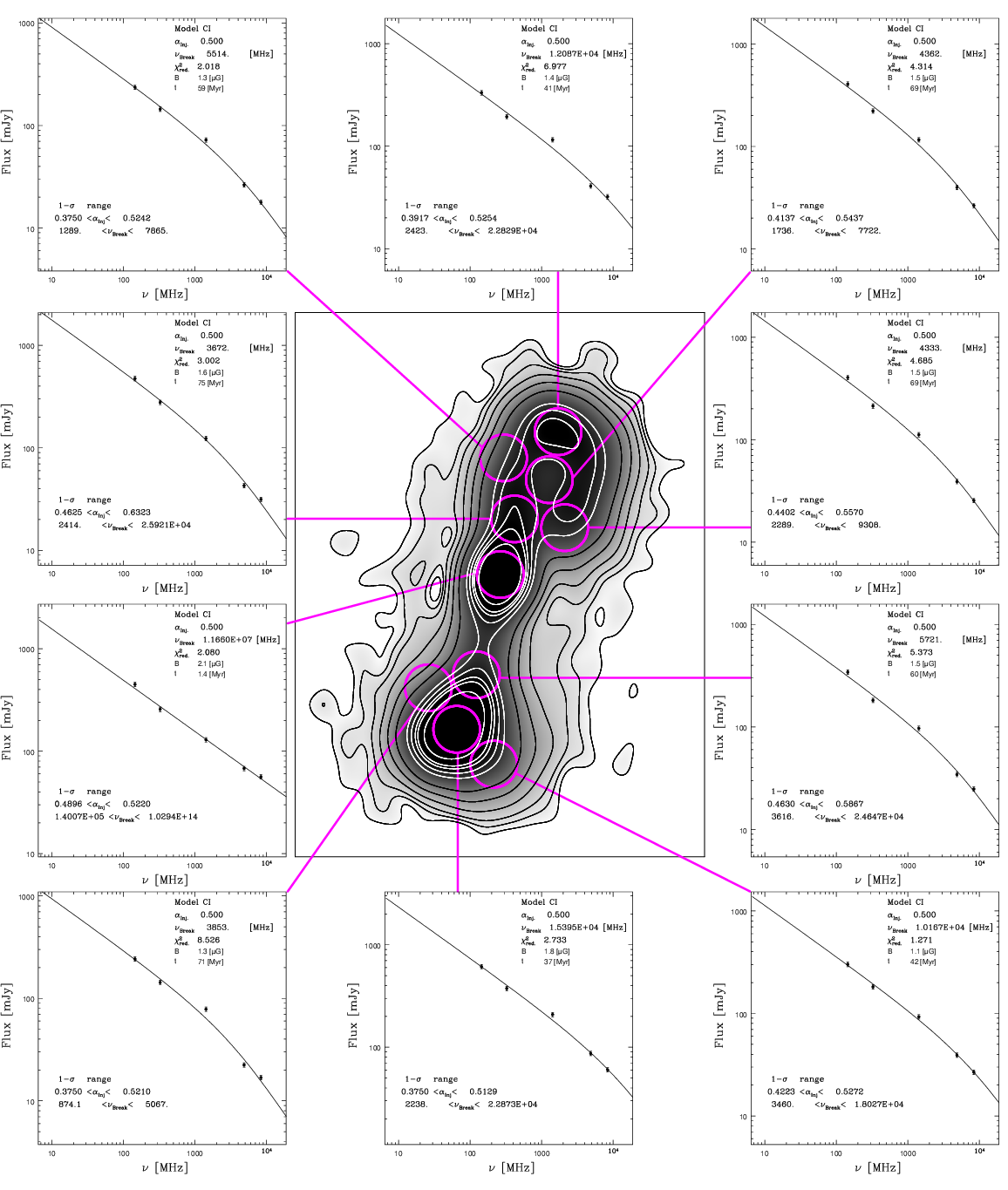}}
\caption{Effelsberg map of 4C\,70.19 from Fig.~\ref{4cm} with the global spectra of selected areas, as marked with the magenta circles (see text for details).}
\label{spixchart}
\end{figure*}

The analyses of the spectral index distribution presented above are performed with the assumption of a constant spectral index in the entire
frequency range analysed in this paper. However, the integrated spectral index presented in Fig.~\ref{globalspix} suggests that spectral indices
can slightly differ between low and high frequency ranges with a distinct change around 1.4\,GHz. The total flux of 4C\,70.19 at 1.43\,GHz is in fact
above the global fit, including uncertainties. This flux was obtained with the use of sensitive VLA C-configuration observations that provided sufficient
resolution for reliable measurements. Therefore, to study the spectral properties of 4C\,70.19 in more detail, we performed
the spectral analysis (both global and in selected regions) of 4C\,70.19 using all radio maps presented in this paper. We note here, however,
that the low-resolution Effelsberg map at 2.67\,GHz was used for modelling of the global spectral index only.
For the analysis of selected regions, all other maps were convolved to the resolution of the Effelsberg map at 8.35\,GHz.
Next, the obtained spectra were fitted with the continuous injection \citep[CI;][]{kardashev62} and the Jaffe-Perola \citep[JP;][]{jaffe73}
models using the {\sc SYNAGE} package \citep{murgia96}. For details of the fitting procedure see \citet{jamrozy08}.
The fitting of the theoretical spectra to our data showed that the CI model worked better (lower values of $\chi^{2}$) than the JP model.
For all fits we used a fixed value of $\alpha_{\rm inj} = 0.5$, as the pre-fitting values were always close to 0.5.
In fact, this value is predicted by shock theory \citep{longair11} and has been also found for regions of hot spots \citep[e.g.][]{carilli91}.
The results of our analyses are presented in Fig.~\ref{globalspixCI} (global spectrum) and Fig.~\ref{spixchart} (selected regions marked with magenta circles).

In our spectral analysis the regions of the diffuse radio emission were not examined, because this emission was detected only in two of our maps,
that is, the LOFAR map at 145\,MHz and the Effelsberg map at 8.35\,GHz (see Sect.~\ref{radiomaps}). The spectral index map between these two
frequencies (left panel of Fig.~\ref{spix}) shows that the values there are between $-0.8$ and $-0.9$.
Our analysis clearly shows, that all spectra, except that for the core region, show a distinct curvature. 
{This means that the synchrotron energy losses of charged particles further from the nucleus are significant
and therefore the plasma there is already aged.} We expect similar behaviour of the spectral
index in areas of the diffuse emission.
For most regions the break frequencies are still below the highest frequency that we analyse in this paper, which is the
8.35\,GHz Effelsberg data. The derived age of the analysed structures varies from around 40\,Myr to 75\,Myr in different regions of 4C\,70.19,
with no visible dependence on the distance from the core.
This might be another argument of the complex morphology of this source.
For the core region we derived an age of 1.4\,Myr only, which points at an ongoing activity of 4C\,70.19.

\section{Summary and conclusions}

We investigated the properties of the radio galaxy 4C\,70.19, which is associated with the
giant elliptical galaxy NGC\,6048. Our main conclusions can be summarised as follows:

\begin{itemize}

\item The sensitive, high-resolution LOFAR map confirms the morphology of the source visible in the high-frequency maps, showing a bent of the
        ``hook-like'' northern jet and a lack of hot spots.
\item In the sensitive LOFAR map at 145\,MHz, and in the single-dish Effelsberg map at 8.35\,GHz, we detect
diffuse radio emission between the central parts of 4C\,70.19 and its radio lobes.
This low-surface-brightness emission was previously not seen in radio maps, most likely due to their low sensitivity to faint diffuse emission.
\item The radio jets seem to rapidly change their orientation when leaving the host galaxy.
        They are also changing their orientation further out, which may be a result of
        tidal interactions with the close companions, especially the galaxy PGC214442, which is present within the elongated infrared halo of NGC\,6048.
\item The propagation of both jets can be explained as a combination of the tidal interactions
        of the host galaxy with its companions and the orbital motion within the IGM.
\item We propose that the elongated brightening along the end of the southern jet is likely a bend, and is
        similar to that of the northern jet, but occurs outwards.
\item The decreasing power of the jets leads to the interaction with the surrounding medium and a consequent change in the direction of propagation,
        manifesting as the shearing of the magnetic field. This shearing changes the orientation of the magnetic field with respect to the jet axis 
	from perpendicular to parallel at the bends.
\item Faraday rotation measurements suggest that the density of the medium around 4C\,70.19 is of the order of 10$^{-3}$\,cm$^{-3}$, 
	which is in close agreement with the transition between the outer parts of the halo of the host galaxy NGC\,6048 and the IGM.
\item We analysed the  spectral index of this source for the first time.
        The global spectral index of 4C\,70.19 ($\alpha$ = -0.61) is relatively flat.
        For different parts of the source, the local spectral index changes by no more than about 0.2.
        This suggests that acceleration and re-acceleration processes should dominate throughout the source,
        as suggested by the spectral index of the faint diffuse emission around the source.
\item The break frequencies calculated for regions in different parts of 4C\,70.19 are at a few GHz. The magnetic fields are roughly uniform throughout
        the source, reaching a value of about 1.5\,$\mu$G, except for the core region and the bright region within the postulated bend of the southern jet.
\item Our analyses suggest that 4C\,70.19 does not show any physical asymmetries, and its de-projected size could be as large as 600\,kpc. 
	We conclude that the diffuse emission around the jets and lobes of 4C\,70.19 are radio plumes expanding outwards, behind the source.
\end{itemize}

The observations and analyses presented in this paper provide a consistent explanation of the intriguing radio morphology of 4C\,70.19; however,
additional data would certainly allow further verification of our findings. These data include sensitive high-resolution observations at both low (GMRT)
and high (VLA) frequencies, deep X-ray imaging of NGC\,6048 and its surroundings, as well as spectroscopic observations of the group galaxies in the vicinity
of 4C\,70.19.

\begin{acknowledgements}
The authors are grateful to thank Alex Krauss for the possibility to perform additional observations with the Effelsberg telescope
and Wojciech Jurusik for substantial help with the LOFAR data reduction. We thank Raffaella Morganti for helpful discussions, as well as 
the anonymous referee for the comments, that helped to improve this paper. 
M.J. is grateful to Matteo Murgia for access to the SYNAGE software. M.J. acknowledges the Polish National Science Centre grant No: 2018/29/B/ST9/01793.
MJH acknowledges support from the UK STFC [ST/V000624/1].
This paper is based (in part) on data obtained with the International LOFAR Telescope (ILT). LOFAR \citep{vanhaarlem13}
is the Low Frequency Array designed and constructed by ASTRON. It has observing, data processing, and data storage facilities in several countries,
that are owned by various parties (each with their own funding sources), and that are collectively operated by the ILT foundation under
a joint scientific policy. The ILT resources have benefited from the following recent major funding sources: CNRS-INSU, Observatoire de Paris and
Universit\'e d'Orl\'eans, France; BMBF, MIWF-NRW, MPG, Germany; Science Foundation Ireland (SFI), Department of Business, Enterprise and Innovation (DBEI),
Ireland; NWO, The Netherlands; The Science and Technology Facilities Council, UK; Ministry of Science and Higher Education, Poland.
The Pan-STARRS1 Surveys (PS1) and the PS1 public science archive have been made possible through contributions by the Institute for Astronomy,
the University of Hawaii, the Pan-STARRS Project Office, the Max-Planck Society and its participating institutes, the Max Planck Institute for Astronomy,
Heidelberg and the Max Planck Institute for Extraterrestrial Physics, Garching, The Johns Hopkins University, Durham University, the University of Edinburgh,
the Queen's University Belfast, the Harvard-Smithsonian Center for Astrophysics, the Las Cumbres Observatory Global Telescope Network Incorporated,
the National Central University of Taiwan, the Space Telescope Science Institute, the National Aeronautics and Space Administration
under Grant No. NNX08AR22G issued through the Planetary Science Division of the NASA Science Mission Directorate, the National Science Foundation
Grant No. AST-1238877, the University of Maryland, Eotvos Lorand University (ELTE), the Los Alamos National Laboratory, and the Gordon
and Betty Moore Foundation.
\end{acknowledgements}

\bibliographystyle{aa} % style aa.bst
\bibliography{myreferences} % all references listed in myreferences.bib

\end{document}